\documentclass[reqno,11pt]{amsart}

\usepackage[toc]{appendix}
\usepackage[utf8]{inputenc}
\usepackage[T1]{fontenc}
\usepackage{microtype} 
\usepackage{graphicx}
\usepackage{hyperref}
\usepackage{amsfonts,amsmath,amsthm,amssymb}
\usepackage[a4paper,bindingoffset=0.5cm,left=2cm,right=2cm,top=2.5cm,bottom=2cm,footskip=.8cm]{geometry}
\usepackage{subcaption}
\usepackage{nth}
\usepackage{bigstrut}
\usepackage[dvipsnames]{xcolor}
\usepackage{colortbl}

\RequirePackage[numbers]{natbib}

\usepackage{tikz}
\usetikzlibrary{automata,positioning}
\usetikzlibrary{decorations.pathreplacing,decorations.markings,shapes.geometric}
\usetikzlibrary{shapes,arrows}
\usetikzlibrary{backgrounds,calc,positioning}

\def \PR {{\mathbb P}}
\def \N {{\mathbb N}}
\def \E {{\mathbb E}}
\newcommand{\Var}[1]{{\mathbb V}{\rm ar}\left(#1\right)}
\def \Cov {{\mathbb C}\mathrm{ov}}
\def \ee {{\rm e}}
\def \eps {{\varepsilon}}
\newcommand{\dd}[1]{{\rm d}#1}

\theoremstyle{theorem}

\newtheorem{example}{Example}
\theoremstyle{plain}

\newtheorem{remark}{Remark}[section]

\newcommand{\vb}{\vspace{3mm}}

\begin{document}

\allowdisplaybreaks

\title[]{\small 
STAFFING FOR MANY-SERVER SYSTEMS\\
FACING NON-STANDARD ARRIVAL PROCESSES}

\author{By M. Heemskerk, M. Mandjes \& B. Mathijsen}

\begin{abstract}

Arrival processes to service systems often display (i)~larger than anticipated fluctuations, (ii)~a time-varying rate, and (iii)~temporal correlation. 
Motivated by this, we 
introduce a specific non-homogeneous Poisson process that incorporates these three features.
The resulting arrival process is fed into an infinite-server system, which is then used as a proxy for its many-server counterpart. 
This leads to a staffing rule based on the square-root staffing principle that acknowledges the three features. 
After a slight rearrangement of servers over the time slots, we succeed to stabilize system performance even under highly varying and strongly correlated conditions. 
We fit the arrival stream model to real data from an emergency department and demonstrate (by simulation) the performance of the novel staffing rule.

\vb \noindent {\sc Keywords.} Queueing; Applied Probability; OR in Health Services.

\vb

\noindent
{\sc Affiliations.} Mariska Heemskerk and Michel Mandjes are with Korteweg-de Vries Institute for Mathematics, University of Amsterdam; Science Park 904, 1098 XH Amsterdam; The Netherlands ({\it email}: {\tt\scriptsize  {j.m.a.heemskerk|m.r.h.mandjes}@uva.nl}). 

Britt Mathijsen is a former PhD candidate at the Department of Mathematics and Computer Science; Eindhoven University of Technology; P.O. Box 513, 5600 MB Eindhoven; The Netherlands. 
({\it email}: {\tt \scriptsize bwjmathijsen@gmail.com}). 
Version: \today.

\vb

\noindent
{\sc Funding.}  The research of the first two authors is funded by the NWO Gravitation Programme N{\sc etworks} (Grant Number 024.002.003). The research of the second author is partly funded by an NWO Top Grant (Grant Number 613.001.352). The research of the third author was funded by an NWO Free Competition Grant (Grant Number 613.001.213).

\end{abstract}

\maketitle

\section{Introduction}
The design of staffing algorithms for service systems has been attracting a great deal of interest ever since Erlang published his first papers.
The delay experienced by the system's users is predominantly caused by the inherent randomness in the arrival stream.
From a managerial point of view, it is natural to address this randomness in such a way that operational costs and customer satisfaction are balanced.
An important complication that recently received a lot of attention is that, as has been observed in various empirical studies, the variance of the arrival stream is larger than the corresponding mean; this phenomenon, called overdispersion, is not captured by standard Poisson processes. 
The challenge that arises is to develop staffing algorithms that are based on more sophisticated, realistic arrival stream models.
See \cite{GKW07,HLW16,JMMW96,LKDJ12,surveyJB} for related work on the design of staffing rules for service systems with overdispersed input.
Such staffing rules have broad application potential, in settings that include call center environments \citep{Aksin2007,BRZ10,Borst2004,GKM03,Whitt99}, cloud computing \citep{Tan2012,Leeuwaarden2016}, and health care delivery \citep{M09}.
 
\noindent {\it Staffing in stochastic service systems.}
Due to the intrinsic stochastic variability of most service systems, they are naturally described by a stochastic model. 
More specifically, queueing models have proven to reveal `good' staffing rules; they determine the number of staff needed to effectively but efficiently cope with the demand imposed on the system.
Evidently, any staffing rule should be such that the average workload brought in per time unit is smaller than the system capacity, to make sure that the delays experienced by patients remain bounded. 
Moreover, certain performance targets are set to guarantee the patients a specific `Quality-of-Service' (QoS) level, which is typically expressed in terms of waiting time.
The objective of the system operator is, on the other hand, to bring the utilization level as close as possible to $1$, so as to control operational cost by efficiently using the resources. 
Staffing rules aim to strike a proper balance between the interests of the patients and the system operator.

The goal that is often strived for is to design a service system in such a way that its patients go into service more or less \textit{immediately upon arrival} \citep{JMMW96,Whitt99}.
Consequently, a commonly chosen service-level agreement (SLA) is to bound the probability of delay by some (typically small) QoS parameter $\eps>0$ \citep{Borst2004,JLZ11, ZLZ12}.
In the typical situation that the arrival rate varies over time, this delay probability is clearly time-dependent too. 
The manager's objective would be to set up a staffing schedule that minimizes the operational costs under the constraint that the delay probability, at any point of time, does not exceed $\eps$. 
The ideal staffing algorithm is one that \textit{stabilizes} the probability of delay over time around some value close to $\eps$, bringing the system in the so-called Quality-and-Efficiency-Driven (QED) regime \citep{HW81}.
Note that, in case there are periods where the algorithm induces a probability of delay significantly smaller than $\eps$, it would mean that (at least locally) `too many' resources are deployed; the variability in the arrival stream is anticipated suboptimally.

\vspace{2mm}

\noindent {\it Realistically modeling the arrival process.} 
Queueing models have been used intensively
to describe and understand congestion phenomena that arise in case of scarce resources.
As a first step in designing a realistic model, it is key to study the arrival process at hand.
We will continue this introduction with a short recap on the properties that should be present in a realistic arrival stream model.

A common assumption in queueing theory is that of Poissonian arrivals, entailing that the mean and variance of the number of arrivals (roughly) match. 
However, it is often observed that service systems face arrival streams that are highly variable (mean $\ll$ variance; overdispersion), while in specific cases systems have to deal with almost deterministic arrivals (variance $\ll$ mean; underdispersion).
As an example of the latter, consider service systems in healthcare with scheduled yet not necessarily punctual arrivals (so that arrival epochs randomly fluctuate around the appointed arrival time), as studied in e.g., \cite{Jouini, KVWC15, KWC17}. 
In such settings clearly some sort of `induced deterministicness' plays a role, in the sense that arrivals are actively being directed to (or away from) the system. 
In this thesis however, we will focus on `undirected' arrival streams only.

For `undirected' arrival streams, overdispersion is a phenomenon commonly found in data.
Examples where one could expect to encounter overdispersed arrivals include a call center of a bank, an insurance company and an emergency department in a hospital; see e.g. \cite{BRZ10, KW14, KW14a, LKDJ12}.
In such settings, arrivals are usually triggered (or inhibited) by occasional events or (un-)favorable circumstances which can cause unforeseen peaks (or dips) on top of the usual daily patterns. 
This so-called `random environment' gives rise to an effect commonly referred to as parameter uncertainty \citep{BRZ10,Whitt99}, which naturally leads to overdispersion. 

Speaking of daily patterns: in nearly all practical applications, the \textit{mean} number of arrivals is not constant over time (e.g. over the course of the day) and follows a predictable pattern. 
It must be noted that the variability that causes overdispersion is of a different nature than the variability induced by nonstationarity.
Nonstationarity can be modeled by a non-homogeneous Poisson proces, replacing the constant arrival rate of a Poisson process by a (deterministic) time-varying one.
However, for non-homogeneous Poisson processes
the mean and variance of the number of arrivals still match, hence such processes fail to capture the entirety of the desired dynamics observed in arrival processes.
Nevertheless, nonstationarity is another important feature of a real-life arrival process \citep{GK91,GKS91,Whitt91} and as such should be incorporated in any realistic arrival stream model as well. 

Besides being overdispersed and having a time-varying rate, a realistic arrival stream might even have dependencies between the numbers of arrivals in disjoint time intervals.
That is to say: it's highly unlikely that the random environment affects the arrival stream in an i.i.d.\ fashion over the different intervals; the effects at hand possibly play a role for a longer period of time.
Indeed, arrival data often exhibits these kinds of dependencies, e.g. in call centers \citep{IERS12, IYES16}. 

\vspace{2mm}

\noindent {\it Existing staffing methods.}
As mentioned, our objective is to develop a staffing rule such that the delay probability is sufficiently low, uniformly over time. 
With this rather stringent service-level requirement in mind it is fairly natural to approximate the system relying on its infinite-server counterpart. 
The famous square-root staffing principle is based on exactly this observation.
In the classical setting (M/G/$\infty$ with arrival rate $R$ and unit-mean service times) it uses that the steady-state number of busy servers is Poisson distributed with mean $R$. 
By asymptotic normality the coefficient of variation (i.e., standard deviation divided by the mean) has the approximate form $1/\sqrt{R}$, and that the steady-state probability of delay in a corresponding finite-server setting with $s$ servers, say $p_s(R)$ can be approximated by 
\[
p_s(R) \approx 1-\Phi\left(\frac{s-R}{\sqrt{R}}\right)
\] 
for large $R$, with $\Phi(\cdot)$ the distribution function of a standard Normal random variable.
For $\beta$ such that $1-\Phi\left(\beta\right) = \eps$ we find for the required number of servers \citep{Whitt92,Whitt93}
\begin{equation} \label{sqst}
s = R + \beta \sqrt{R}.
\end{equation} 
Note that Eqn.\ \eqref{sqst} has an appealing interpretation: the number of servers $s$ should evidently be taken larger than the expected workload $R$, with the extra term $\beta\sqrt{R}$ (`uncertainty hedge') being of the same order as the natural load fluctuations of the workload process.
Refinements of order smaller than $\sqrt{R}$ are explored in e.g. \cite{JLZ11, MJLZ17,ZLZ12}. 

Although the excess probability $\PR(M>s)$ corresponding to the infinite-server system (with $M$ denoting the stationary number of busy servers in this M/G/$\infty$ queue) is likely to be smaller than 
the probability of delay in its finite-server counterpart,
still square-root staffing rules have shown to give accurate results \citep{Borst2004,JLZ11}. 
This can be explained by the fact that as $R$ grows large, the hedge $\beta \sqrt{R}$ prevents congestion more and more effectively.

So far we discussed the situation of a constant Poissonian arrival rate.
For large-scale systems the predominant assumption in the literature is that patients arrive according to a time-varying Poissonian arrival rate. 
Staffing algorithms for {non-homogeneous} Poisson processes (NHPPs) have been studied for several decades.

If the arrival process is an NHPP with nonstationary arrival rate $\lambda(\cdot)$, then the number of arrivals $N(s,t)$ in the interval $[s,t)$, with $s<t$, is Poisson distributed with parameter
\[\int_{s}^t \lambda(r)\,{\rm d}r.\]
Note that such a non-homogeneous arrival process not yet exhibits overdispersion ($\E \,N(s,t) = \Var {N(s,t)}$). 
For the resulting model M/G/$\infty$-based staffing rules cannot be applied directly, but various approaches have been proposed. 

In a first approach, the nonstationarity is essentially ignored: one uses a {simple stationary approximation} (SSA), based on a stationary model in which the arrival rate is chosen equal to the long-run average \citep{RO79}.
This method performs poorly in most scenarios \citep{GKS91}, for example when the actual rate is slowly varying with respect to the service time or when the relative amplitude (level of nonstationarity) of the rate is larger than 10\%.
A second approach is the {pointwise stationary approximation} (PSA) \citep{GK91,GKS91,Whitt91}, which considers the system at time $t$ as if it has dealt with an arrival rate $\lambda(t)$ with $s_t$ servers from the start (i.e., assuming steady state), thus ignoring nonstationarity in a different way. 
This method works well in settings where the arrival rate changes sufficiently slowly \citep{GK91,Whitt91}, so it covers scenarios on the other side of the spectrum. 

As a comprimise between the two extremes, \cite{Whitt91} suggests the {average stationary approximation} (ASA) that generalizes both SSA and PSA, replacing the arrival rate at time $t$ with 
\[\bar{\lambda}_{t} = \frac{\E\, S}{a} \int_{t - a/\E\, S}^t \lambda(r)\, \mathrm{d}r\] 
for some positive constant $a$ and mean service time $\E\, S$.
An alternative to this was proposed by \cite{JMMW96}, saying that one could replace $R$ in the staffing formula by $m_{\infty}(t)$, the expected `offered load'  in an infinite-server system with time-dependent arrival rate $\lambda(t)$ at time $t$:
\[
m_{\infty}(t)= \E \int_{t-S}^t \lambda(r)\, \mathrm{d}r,
\]
where $S$ denotes the service time.
They showed that this method stabilizes the probability of delay close to some target value at all times, independently of the arrival rate being slowly varying or not. 
Importantly, in \cite{JMMW96} asymptotic normality was used to arrive at the approximation, hence their method follows the tradition of the square-root staffing procedure described above.

As mentioned above, NHPP models fail to exhibit overdispersion, a phenomenon observed across various types of service systems; see e.g., \cite{CH01, JK01, KW14, Robbins2010, SHM09}.
The parameter uncertainty underlying overdispersion potentially jeopardizes the effectiveness of the square-root rule, typically leading to overoptimistic staffing algorithms. 
This complication was brought forward in many studies, e.g.\ in \cite{ADE04,BRZ10,GKM03,Grassmann88,GLT10,JK01,KW14,MJLZ17,Mehrotra2010,SHM09,Zan2012}.
Different methods were proposed to overcome this issue;
typically a \textit{Poisson mixture} is used to model parameter uncertainty.
That is, the deterministic Poissonian arrival rate is replaced by a sampled one; see \cite{Grassmann88,Whitt99, CH01,JK01,BZ09,M09,KAW15,MJLZ17} for examples.

Relatively little attention has been paid to staffing rules in the context of arrival processes in which the numbers of arrivals in disjoint intervals are dependent. 

\vspace{2mm}

\noindent {\it Contributions and organization.}
The contributions of this paper are twofold. 
In the first place, we present a flexible model for the arrival process, based on \cite{HLM17}, that can deal with (any level of) overdispersion, nonstationarity and dependencies between arrivals of consecutive time slots. 
The challenge being to come up with a model that remains of practical use/computationally tractable, we believe that the model proposed here is among the simplest models with these three properties.
Moreover, fitting data to our model is a relatively straightforward task.
The model is presented in Section \ref{MD} 

In the second place, we develop staffing rules meeting the criteria mentioned in the introduction, to go with this comprehensive yet simple model for the arrival stream.
It requires low computational cost to determine staffing prescriptions based on these rules.  
In Section \ref{SR} we present the new staffing rule.
Subsequently, in Section \ref{intro.data} we present a case study based on a healthcare-related data set to show that the rule succeeds to stabilize the delay probability around the targeted $\eps$. 
The observations here lead to a much improved version of the staffing rule that was introduced in Section \ref{SR}.
This concludes Section \ref{sec2}.

In the rest of the paper we work out the details necessary for implementation and further asses the performance of the proposed staffing rules.
Section \ref{gap} presents straightforward statistical procedures to estimate the modeling parameters. 
We perform the estimation procedure both for a real data set from an emergency department, and for a stylized example. 
In Section \ref{per} we perform extensive experiments to assess the performance of our staffing rule in settings with overdispersion, a time-varying arrival rate, and temporal correlation.  
Here, we incorporate impatience into the model (as in reality, customers might abandon the system before their service begins), in order to analyze how this affects the performance. 
Finally, Section \ref{CD} concludes the paper.

\section{Model and staffing rule} \label{sec2}

In this section we first present our arrival stream model meeting the criteria mentioned in the introduction (overdispersion, time-varying rate, correlation between disjoint time intervals). 
Our model is arguably the simplest among all models satisfying these requirements. 
It is relatively compact and only requires a few input parameters. 
We then introduce a suitable staffing rule to match such arrival streams.
It is new compared to the earlier described methods in the introduction, as it combines all three features of realistic arrival processes while still using the concept of square-root staffing, where the mean under the square-root is replaced by the variance of the number of customers in the approximative infinite-server system.
We conclude the section by an illustrative case study, in which we demonstrate the procedure and its performance. 

\subsection{Model description}\label{MD}
The model we consider could be termed a \textit{mixed} $M_t/G/s_t$ queue with infinite waiting room.
We systematically introduce the components of the model, starting with the arrival process. 

\vspace{2mm}

\noindent {\it Arrival process}. 
In our setup the arrival process is a {Cox process}, i.e., a time-dependent Poisson process with {random} arrival rate. 
At time $t$, the arrival rate is $\Lambda(t)\geqslant 0$. 
This $\Lambda(t)$ consists of a \textit{deterministic trend} $\lambda(t)$ (capturing the daily pattern), which is inflated by a \textit{stochastic busyness factor} that incorporates the desired overdispersion and temporal correlation. 
As a consequence, the model proposed possesses the three desired properties. 

More specifically, the arrival rate is built up as follows.
Following common practice, we assume that $\lambda(t)$ is piecewise constant on time intervals of fixed size $\Delta$.
For $t \in[j\Delta,(j+1)\Delta)$  we can therefore write  $\lambda(t) = \lambda_j$. 
We introduce a sequence of random variables $W\equiv \{W_j\}_{j\in\mathbb{Z}}$, which are independent and distributed as a random variable $W \geqslant 0$; we normalize them such that  $\E W=1$, and assume $\Var W < \infty$. 
The busyness factor of slot $j$ is affected by the current value of the $W$ process ($W_j$, that is), but also by the previous $I$ values ($W_{j-I}$ up to $W_{j-1}$). 
The parameter $I\in \N$  reflects the amount of dependence between the stochastic arrival rates pertaining to consecutive disjoint slots.  
Let $N$ be the total number of time slots of size $\Delta$ in the considered time frame, i.e. $N=24$ when $\Delta=1$ hour and the considered time frame is a day. 
The level of dependence from previous values of the $W$ process is dealt with in an autoregressive way, with parameter $\alpha\in(0,1).$
Concretely, this means that for $t \in [j\Delta,(j+1)\Delta)$ the stochastic arrival rate is given by
\begin{equation} \label{arr}
\Lambda(t)= \lambda_j \cdot \Big(c_{\alpha}\sum_{\ell = 0}^I \alpha^\ell W_{j-\ell} \Big);
\end{equation}
here $c_{\alpha} :=(1-\alpha)/(1-\alpha^{I+1})$ is a normalizing constant that ensures that the busyness factor has mean 1:
\[
\E\left(\sum_{\ell = 0}^I \alpha^\ell W_{j-\ell}\right) = \frac{1-\alpha^{I+1}}{1-\alpha}=\frac{1}{c_\alpha}.
\]
It means that the busyness factor gets a new value every $\Delta$ time units; thus, $\Delta^{-1}$ can be regarded as the \textit{sampling frequency}. 
Note that the process $W$ is {\it not} observable; as we show later, in our staffing formula we just need to know $\Var{W}$. 

The values $\lambda_j$ reflect the mean arrival rates during the individual time slots. 
When assuming periodicity in the data (e.g., daily and weekly patterns), one can estimate these values in a straightforward way from historic data. 
The value of $\Delta$ is situation-dependent; one often picks $5$, $10$ or $15$ minutes. 
This leaves us with estimating $\alpha$, $I$, and $\Var W$. 
The procedure we followed is that we use standard least-squares tools to estimate $\alpha$ and $\Var W$ for given $I$; this we do for multiple values of $I$, so as to select an `optimal' $I$. 
We elaborate on these estimation issues in Section \ref{gap}. 

\vspace{2mm} 

\noindent {\it Service times.} 
The patients' service times are independent and identically distributed samples from a general non-negative distribution; we denote the underlying random variable by $S$, and write $P(t) := \PR(S>t)$. 
In the numerical experiments in Sections \ref{intro.data} and \ref{per} we focus on the case of exponentially distributed service times (with mean $\mu^{-1}$), but in the staffing rule one could pick in principle any distribution.

\vspace{2mm} 

\noindent {\it Number of servers.} 
At time $t$, the number of servers is $s_t$. 
The value of $s_t$ is as determined in Section \ref{SR}. 
We assume that services are always completed, even if $s_t$ drops to a value that is insufficient to serve all patients present; as this assumption is fairly natural in practice this is the way the system dynamics will be modeled in the simulation experiment.

\subsection{Staffing rule}\label{SR}
The staffing rule we propose is essentially an adaptation of the classical square-root staffing rule in Eqn.\ \eqref{sqst}: for some constant $\beta>0$,
\begin{equation}\label{superrule}
s_t=m_\infty(t) + \beta \sqrt{v_\infty(t)};
\end{equation}
here $m_{\infty}(t)$ and $v_\infty(t)$ are the mean queue length and variance of the mixed $M_t/G/\infty$ counterpart of the mixed $M_t/G/s_t$ system introduced in Section \ref{MD}. 
Note that given an overdispersed arrival stream, the term $\beta \sqrt{v_\infty(t)}$ (the hedge) is larger than in the classical SRS rule, where it would equal $\beta \sqrt{m_\infty(t)}$.

The use of such a rule is justified by asymptotic normality, which is backed by the results in \cite{HLM17}.
The initial choice for the constant $\beta$ is (with $\Phi(\cdot)$ the normal CDF):
\begin{equation}\label{constantnormal}
\beta = \Phi^{-1}(\varepsilon).
\end{equation}
It is expected that this choice is not optimal, given the approximative nature of the procedure.
In fact, $\beta$ is always smaller than optimal, since the actual number of customers in a finite-server system will be higher than predicted by an infinite-server proxy (where each customer is served immediately upon arrival and hence can leave without waiting).
The idea is to slightly tweak the value $\beta$ in order to more closely attain the desired service level (i.e., $\PR(\text{delay}) < \eps$). 
More importantly, irrespective of the level the shape of $s_t$ as a function of time should ensure a delay probability that is relatively flat over time.
This depends mostly on the shape of $m_{\infty}(t)$ and $v_\infty(t)$.

Hence, the next step is to determine expressions for $m_{\infty}(t)$ and $v_\infty(t)$.
Following the approach of \cite{HLM17}, we deduce that the queue-length process of an infinite-server queue fed by a Cox process arrival process with arrival rate $\Lambda(t)$ is again a Cox process. 
More specifically, the distribution of the number of patients at time $t$ is Poisson with random parameter
\begin{equation}
R(t) =\int_{-\infty}^t \Lambda(s)\,\PR(S>t-s)\, \dd s.
\end{equation}
We thus obtain that
\begin{equation} \label{molmean}
m_{\infty}(t) = \E R(t) = \E \left[\int_{0}^{\infty} \Lambda(t-s)\,\PR(S>s)\, \dd s\right]
\end{equation}
and, by the law of total variance (conditioning on $\Lambda(s)$, for $s\in(-\infty,t]$),
\begin{align} \label{molvar}
v_{\infty}(t) = \Var {R(t)} &= \E [\Var {M(t)} \mid \Lambda(\cdot))] + \Var {\E [ M(t) \mid \Lambda(\cdot)]} \nonumber \\
&= m_{\infty}(t) + \Var {\int_{0}^{\infty} \Lambda(t-s)\,\PR(S>s)\, \dd s}.
\end{align}
As an aside, note that indeed $m_{\infty}(t)\leqslant v_\infty(t)$, which reflects the overdispersion that we introduced. 
These expressions can easily be simplified using the observation that $\Lambda(t)$ is piecewise constant.
For the case that $S$ is exponentially distributed, $\mu_{\infty}(t)$ and $v_\infty(t)$ can be evaluated in closed form in $t=n\Delta$ with $n\in \N$; see Appendix \ref{app1} .

\subsection{Case study: MOL staffing for overdispersed hospital arrival data} \label{intro.data}
We continue by illustrating our approach and its performance in a case study. 
The data set used was provided by the SEElab and originates from the emergency department  (ED) of an Israeli hospital. 
It contains $5$-minute resolution arrival counts of a $4$-year time period ($1999-2003$), which covers a total of $1569$ days.
The average arrival volume per day, exceeding $300$ arrivals, is sufficienty large and the mean length of stay (LOS) is almost $2$ hours.

\begin{figure}
\centering

\vspace*{20pt}

\textbf{Arrivals}\par\medskip
	\begin{subfigure}[b]{0.49\textwidth}
	\includegraphics[width=\textwidth]{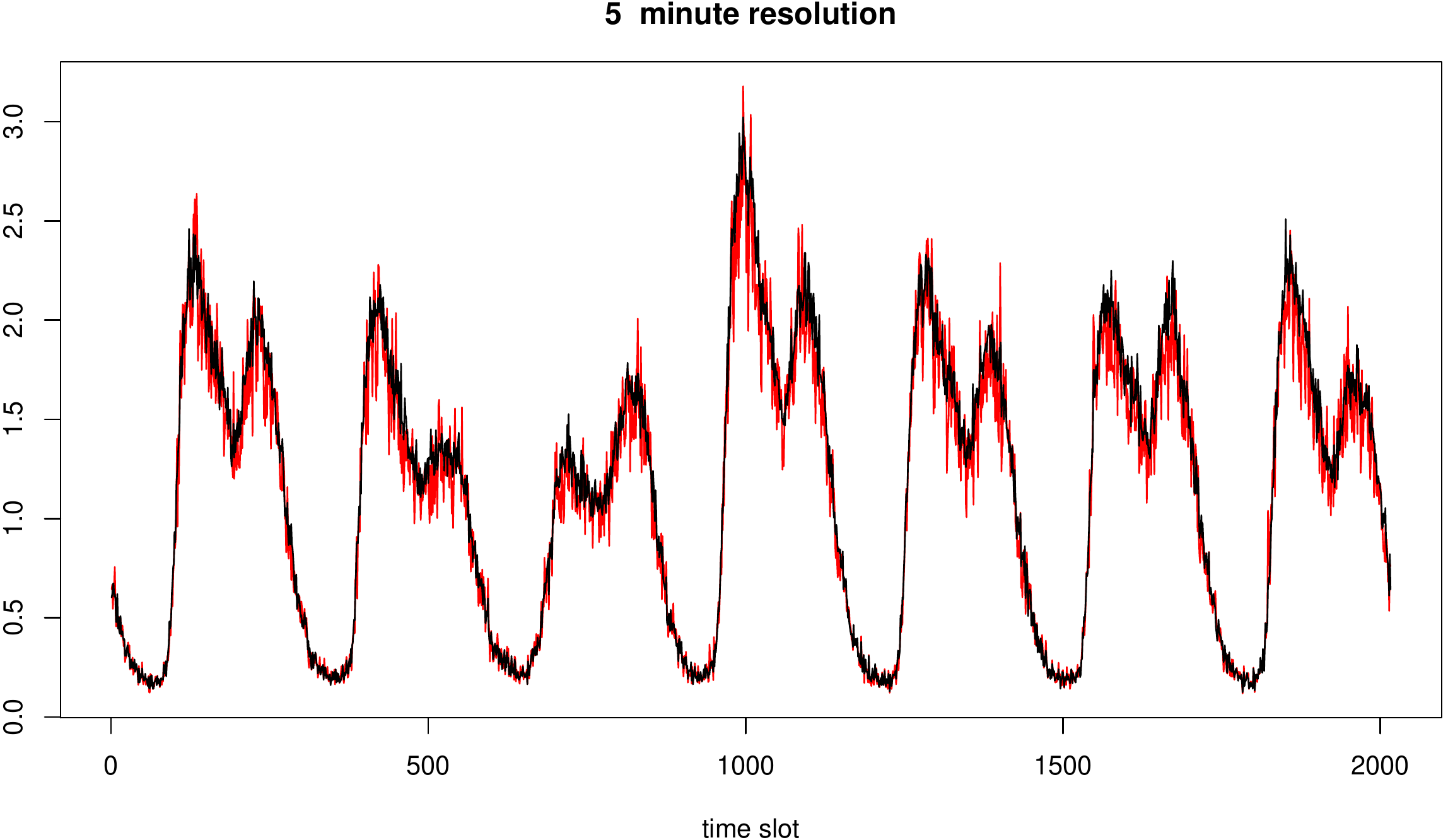}
	\end{subfigure}
\hfill
	\begin{subfigure}[b]{0.49\textwidth}
	\includegraphics[width=\textwidth]{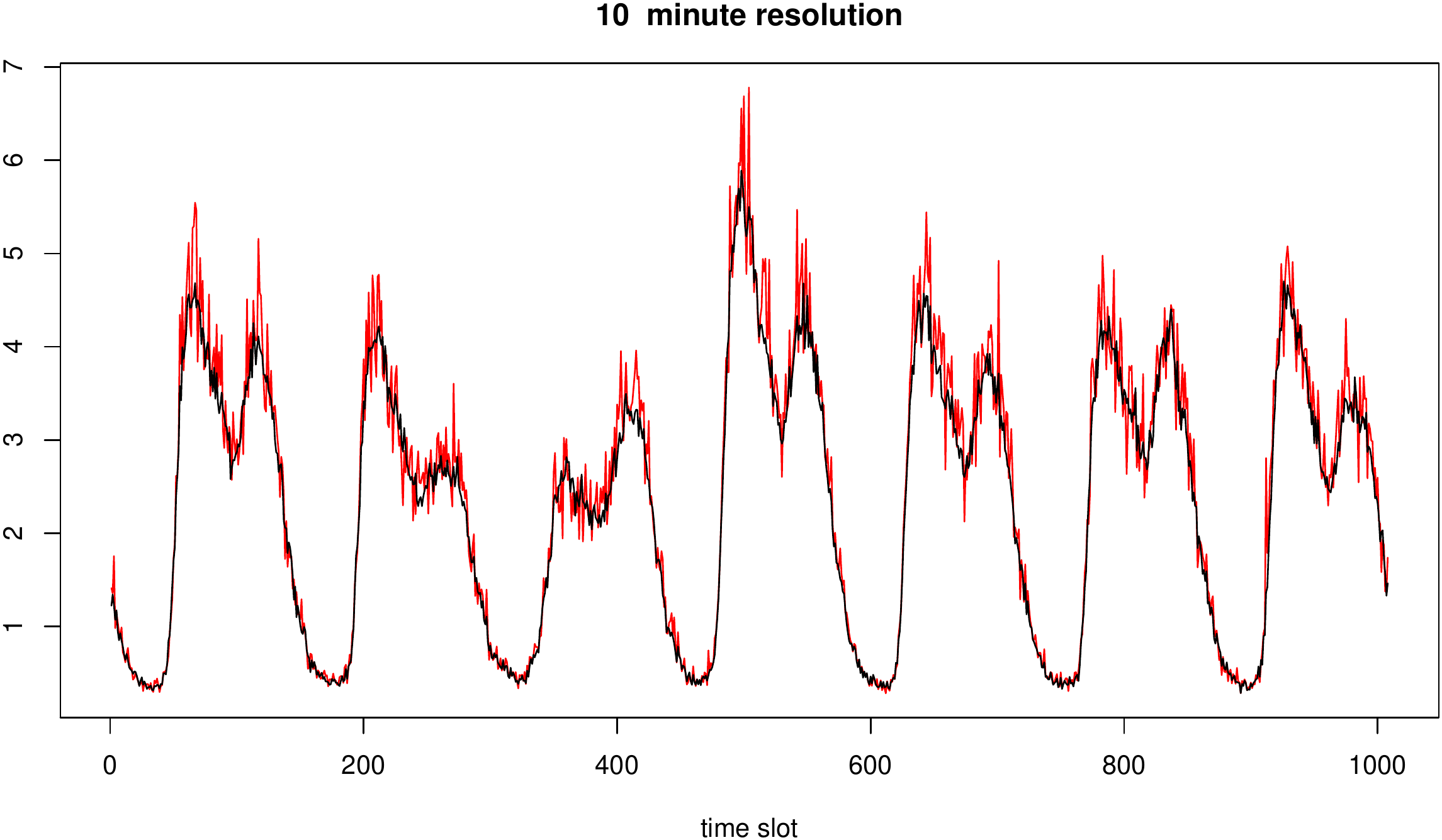}
	\end{subfigure}
	
\vspace*{20pt}

	\begin{subfigure}[b]{0.49\textwidth}
	\includegraphics[width=\textwidth]{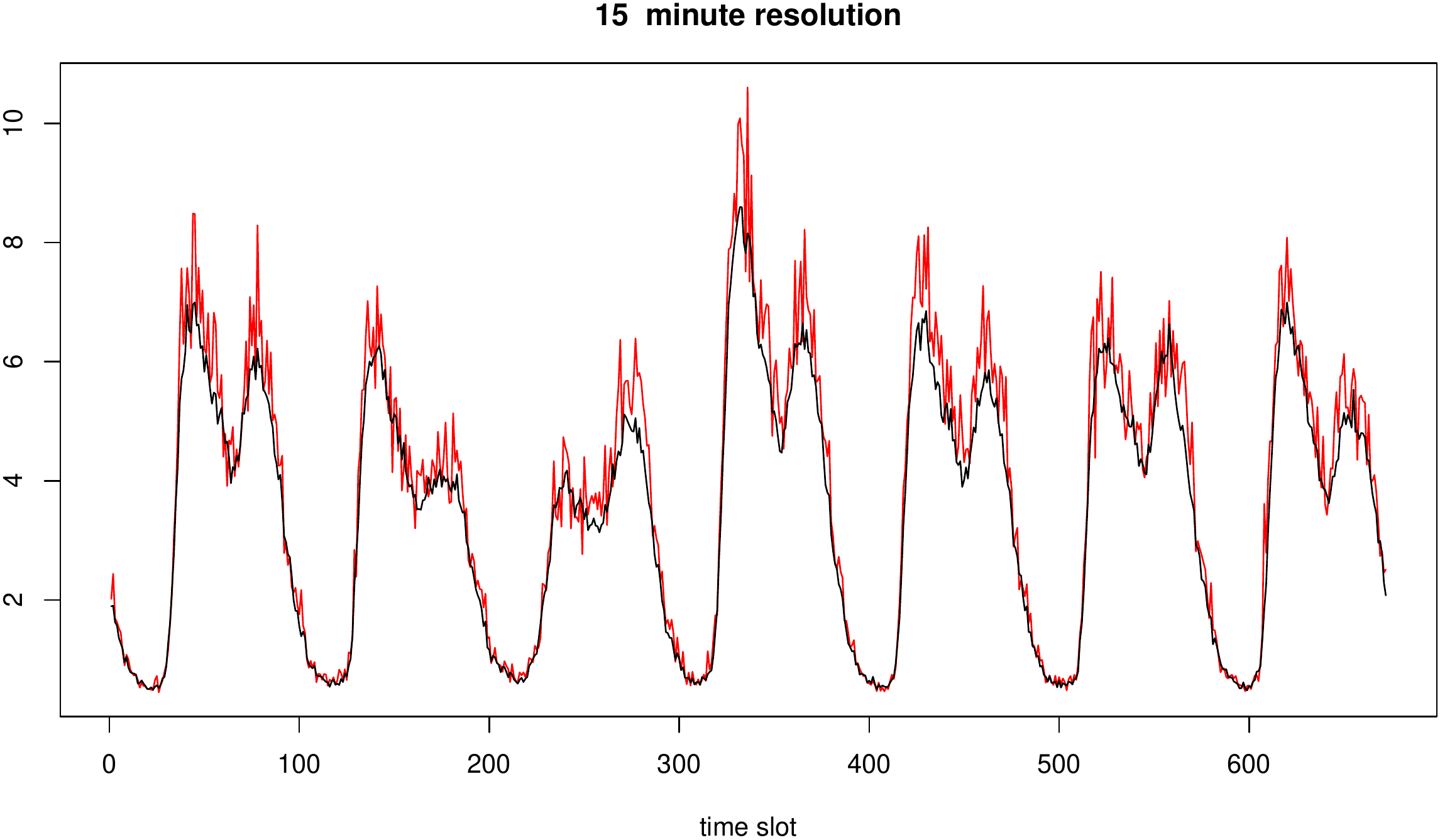}
	\end{subfigure}
\hfill
	\begin{subfigure}[b]{0.49\textwidth}
	\includegraphics[width=\textwidth]{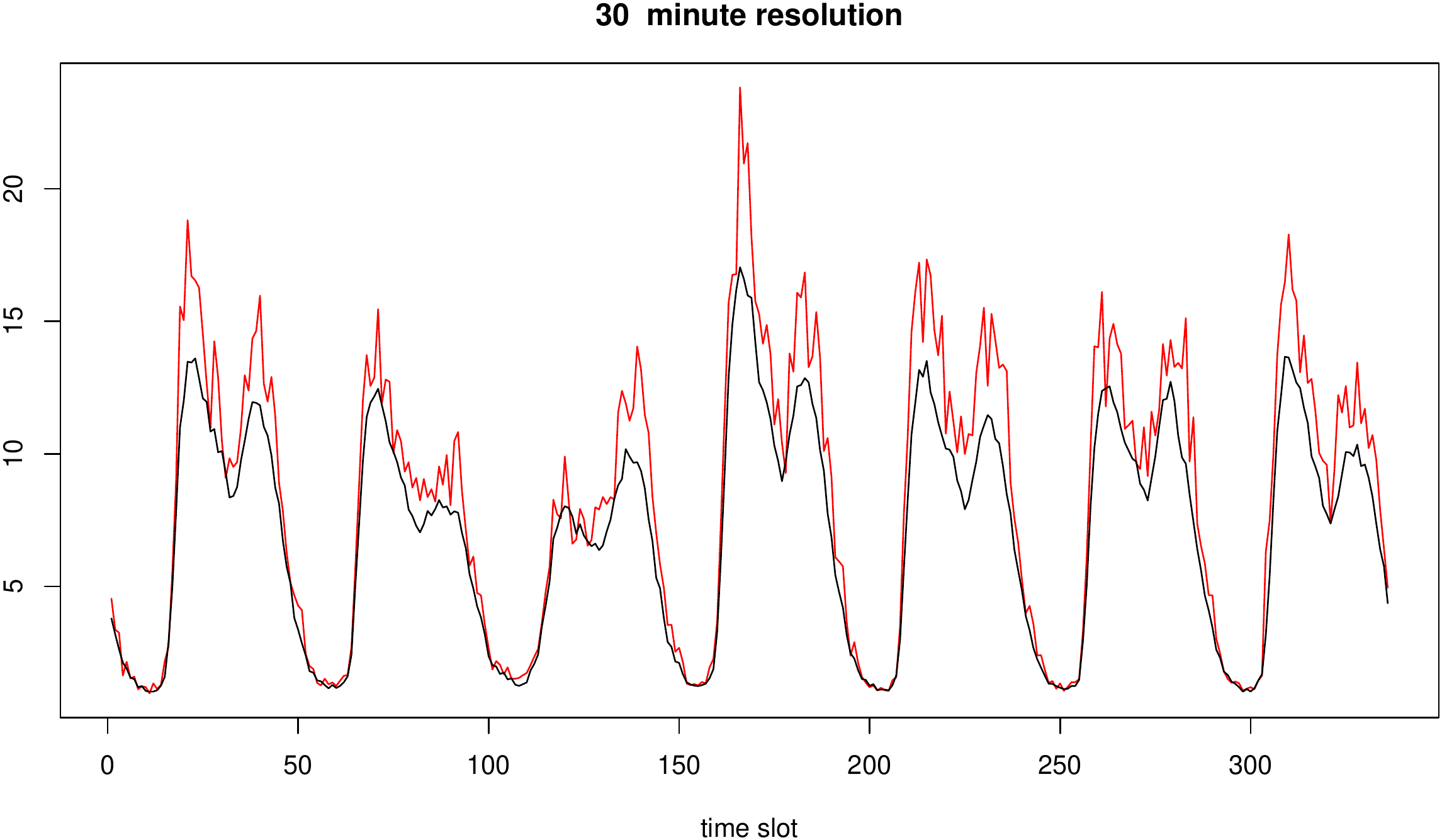}
	\end{subfigure}
	
\includegraphics[scale=0.35]{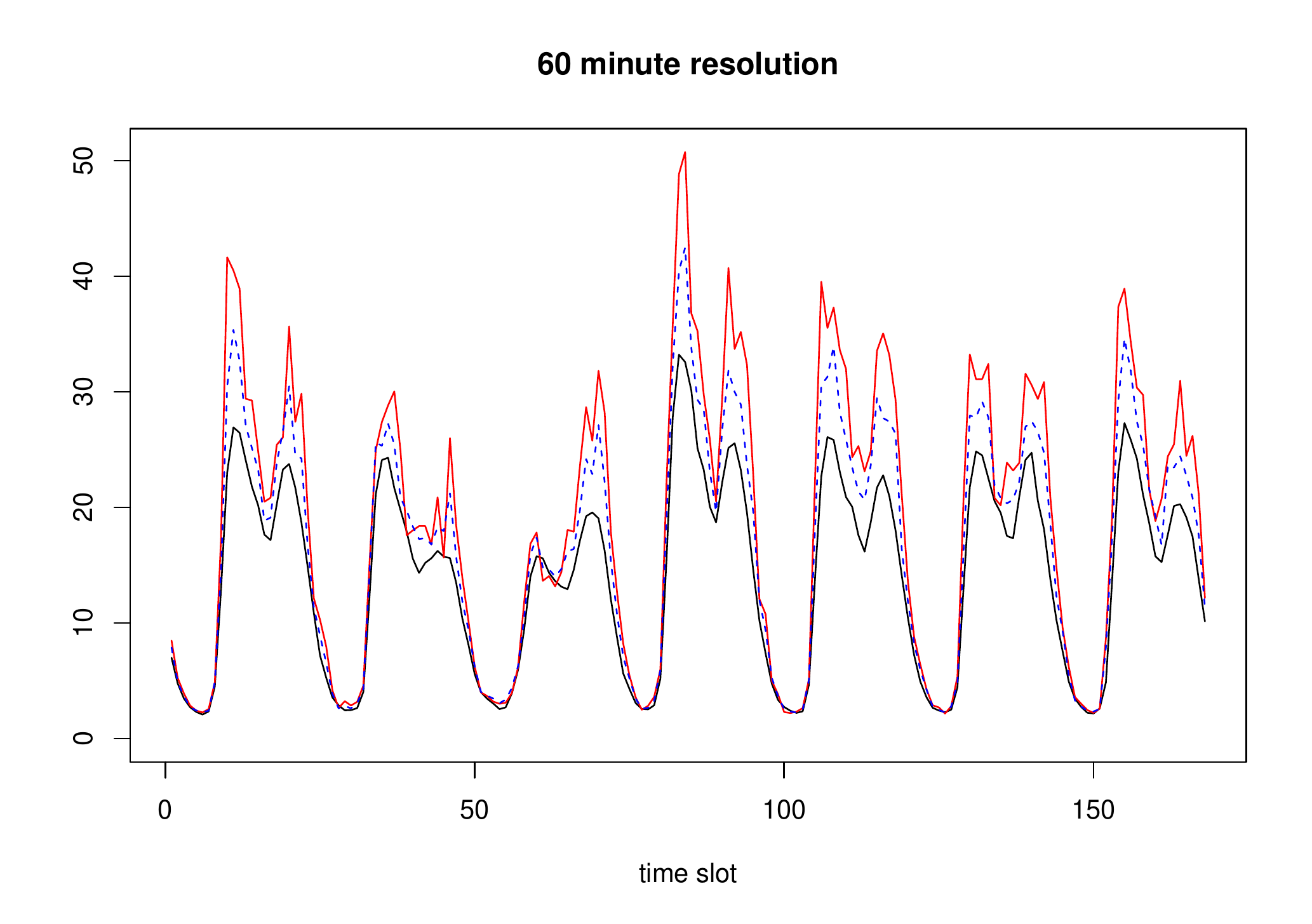}			
\caption[justification=raggedright]{\label{fig:week_res} 
Mean (black) and variance (red) of the number of arrivals in each time slot for various resolutions. 
The week starts at Thursday, as the first day arrival data was recorded was a Thursday.
The dashed blue line in the 60 minute resolution plot shows the sum of the variances in the corresponding 30 minute time slots.
The gap between the dashed blue line and the red line indicate the presence of nonnegative correlation between the number of arrivals in consecutive time slots.
}
\end{figure}

We aggregate different weekdays separately,
which implies that we have $N=224$ observations for each day of the week (see Table \ref{table:stats}). 
Figure \ref{fig:week_res} presents the sample mean and variance of the number of arrivals per slot for a 5, 10, 15, 30 and 60 minute resolution, based on these $224$ observations.
When considering the hourly mean arrival rates (same timescale as LOS) for each hour of the week it is fairly variable: with a time average of $14$, its minimum is $2.1$ and its maximum $33.2$. 
In conclusion, the level of nonstationarity is high and the rate is rapidly changing with respect to the LOS.
Note that this means that both of the methods mentioned in the introduction, SSA and PSA, would not be accurate.

\begin{table}[h]
\centering
\begin{tabular}{|r|r|}
\hline
\hline
\textbf{Start day} & 1-4-1999 \\
\textbf{End day} & 17-7-2003 \\
\textbf{Total \# days} & 1569   \\
\textbf{Total \# weeks}  & 224 \\
\textbf{\# Arrivals per day}  & 324.59  \\
\textbf{Mean LOS (min.)}  & 109  \\
\textbf{St. dev. LOS (min.)}  & 114  \\
\hline\hline
\end{tabular}%
\caption{Summary statistics of the hospital ED.}
\label{table:stats}
\end{table}

An observation from Figure \ref{fig:week_res} is that different weekdays indeed show different patterns in the arrival stream and also the level of overdispersion and nonstationarity is visibly different although data has already been averaged over 224 samples.
Note for example that Sunday (in Figure \ref{fig:week_res} the \nth{4} day) is an exceptionally busy day with high peaks in the mean and variance, in contrast to Friday and Saturday (i.e., Israeli weekend).

Furthermore, observe that the (positive) difference between the mean and the variance increases with the chosen resolution, but only when choosing a resolution larger than 30 minutes overdispersion becomes more apparent in the plots.
Comparing the values corresponding to the different subfigures in Figure \ref{fig:week_res}, it shows that the growth in the variance is still roughly linear.
In fact, our model predicts what we observe in Figure \ref{fig:week_res}: the variance in the number of arrivals (just like the mean) grows roughly linear in the length of the time slot, but due to the presence of (nonnegative) correlation between rates in consecutive time slots, an extra term should be added to the variance when aggregating data from smaller time slots.
On top of this visualization of the sample means and variances, we can also compute the empirical covariance matrix to quantify the correlations between the arrival counts in all different time slots, for each of the resolutions.

\vspace{2mm}

When fitting the model to the arrival data we find that choosing the parameters $\alpha$, $I$ and $\Var W$ differently for different weekdays significantly improves the fit.
As this modeling decision also affects the staffing rule and the subsequent performance analysis, we decided to simplify reality and examine only isolated Sundays in the rest of this case study.
Note that consequently we pretend that Sundays succeed one another, so that time slots around midnight are correlated in the model although in reality there is a week of (ignored) events in between.
It is expected that the error resulting from this simplification is small as the arrival volume is small around midnight, and even smaller for small values of $I$.
   
In the rest of this subsection we will present our rule's performance when staffing a multi-server system where the arrival stream is taken from the data set, restricted to Sundays. 
In Section \ref{per} we present a systematic evaluation using stylized input.

Given $\Delta = 1$ hour and $I = 10$ (where we intentionally pick a large value of $I$ to be on the safe side), we find by the statistical inference procedure that will be described in Section \ref{gap} (cf.\ Table \ref{table:gain}) that 
$\alpha =  0.81$ 
and $\Var W =  0.11$ 
are the best fit for the data.
As this will be the input for the simulation, we expect that using these parameters for the staffing rule will give the best result.
To check this, we will also generate the delay probabilities if we plug in different parameter settings in the staffing rule, i.e., $I=0,1,5$ (with corresponding $\alpha$ and $\Var W$ as given in Table \ref{table:gainSun}).
As in Table \ref{table:stats}, the mean length of stay is $109$ minutes, so the hourly service rate to be used in the simulation is $\mu=60/109\approx 0.55$.

\begin{figure}[ht!]
\centering
\includegraphics[scale=0.6]{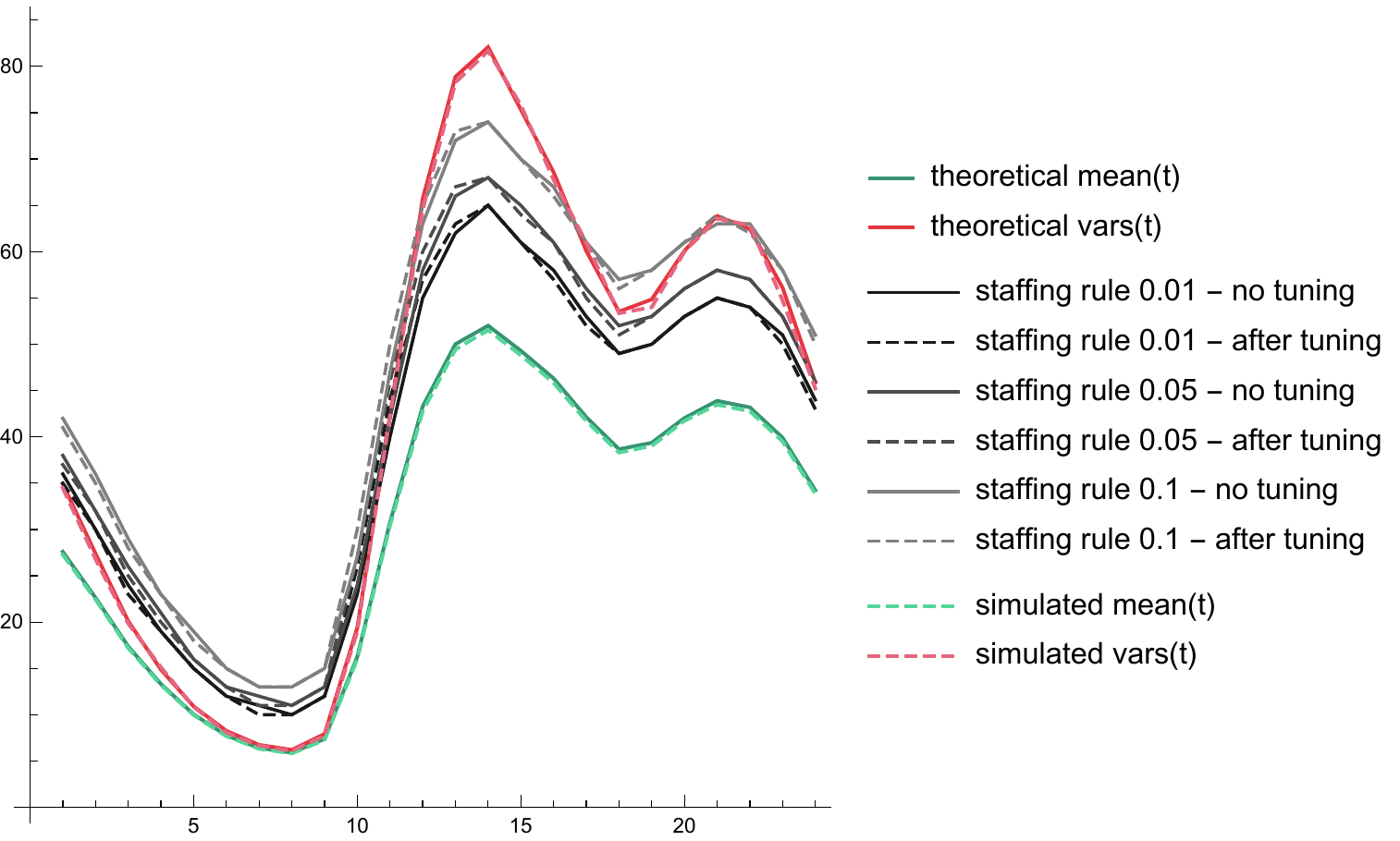}
\caption{\label{fig:results0}Mean, variance and staffing rules given different service-level agreements.}
\end{figure}

Figure \ref{fig:results0} shows the empirical mean and variance obtained from the simulation, which determines the probability distribution of the number of patients per time slot.
Observe that the theoretical values at the end of the time slot, as given in Eqns.\ \eqref{eqminfty} and \eqref{eqvinfty}, more or less coincide with these empirical values; see Figure \ref{fig:results0}.
The prescribed number of servers in time slot $n$ depends on the service level that was set and is chosen according to the staffing rule in Eqn.\ \eqref{superrule}, with $t=n\Delta$ and $\beta$ as in Eqn.\ \eqref{constantnormal}.
We compare $\varepsilon \in \{0.01,0.05,0.1\}$; see the solid gray lines in Figure \ref{fig:results0}.

\begin{figure}[ht!]
\centering
\includegraphics[scale=0.597]{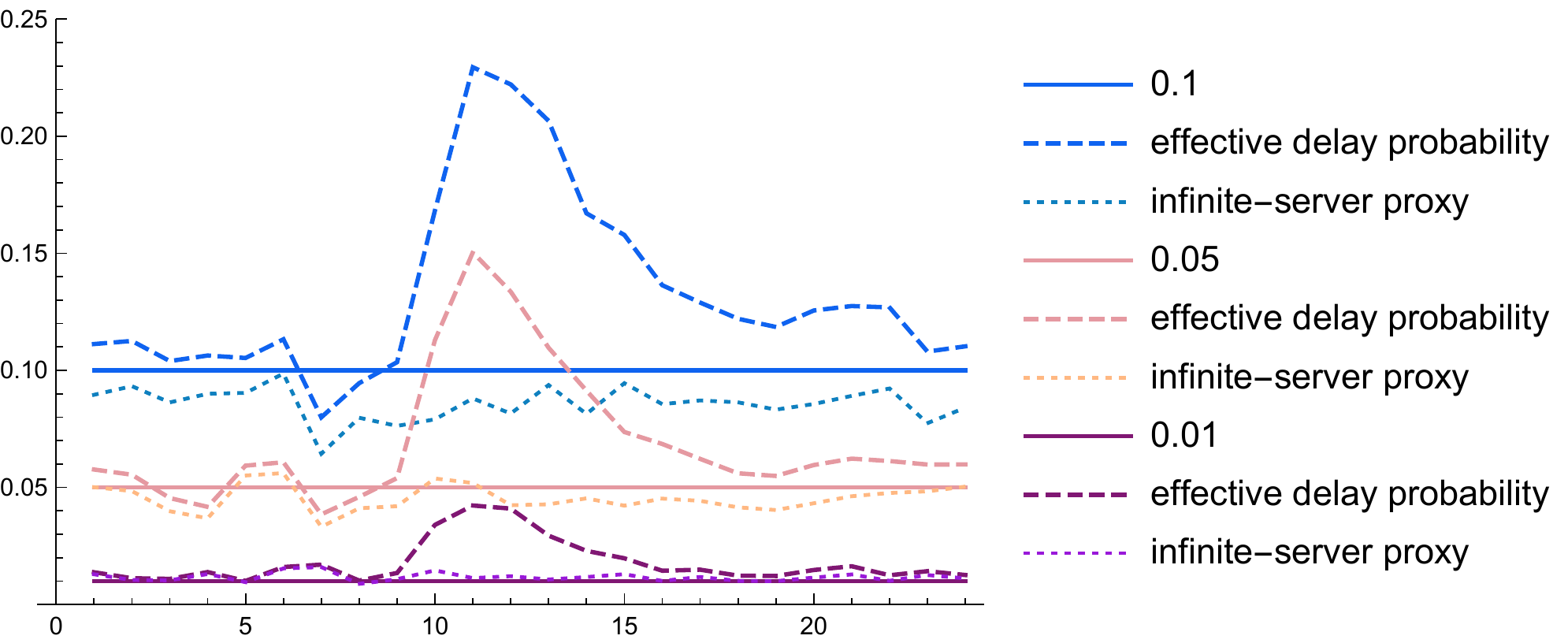}
\caption{\label{fig:results1}Exceedance vs delay probability for different service levels. Each service level is designated by a different color, where the dashed line describes the effective delay probability for the finite-server system under study and the dotted line describes the exceedance probability for the infinite-server proxy.}
\end{figure}

Next, the empirical probability of exceeding the staffing level in an infinite-server system is computed using the simulation results.
As the staffing rule is based on analysis of infinite-server systems, it can be expected that this probability behaves well: asymptotic normality predicts that it should be close to the required level $\varepsilon$ in every time slot, which implicitly says that the service level should be more or less stable.
However, Figure \ref{fig:results1} shows that the exceedance probability sometimes crosses the required service level and does not follow a smooth straight line.

This can partly be explained by rounding errors (note that the `tooths' in the lines are often caused by a difference of $1$ server), and moreover it is noted that asymptotic results in the end are just approximations.
All in all the performance of this very straightforward and easy-to-use staffing rule is satisfying.
But importantly, the infinite-server results can of course only be used as a proxy.
The actual delay probabilities from the finite-server setting (where the staffing rule dictates the number of patients that can be served at a time) will be significantly larger due to queueing caused by the waiting patients. 
If this queueing bias would result in a uniform shift upwards, the staffing rule would still prove perfectly useful, as we can easily tune the delay probability down by tweaking $\beta$.
Unfortunately this is not the case; Figure \ref{fig:results1} shows a heavy spike around noon, so the system is locally performing unacceptably poorly.  
Note that in the (finite-server) setting with abandonments the performance would certainly be better, depending on the abandonment rate.
Now, instead of going immediately into service as in the infinite-server setting, all customers initiate an exponential clock (with a rate that might even be comparable to the service rate) right upon arrival, for potential abandonment of the system.
The infinite-server proxy is way more accurate in such a setting. 
In Section \ref{per} we will consider this adaptation, but for now we try to further improve the staffing rule for the basic setting (i.e., the setting without abandonments).

Note that, although for most of the day the delay probability seems rather stable, around noon it takes on values twice the targeted service level.
Comparing Figure \ref{fig:results0}, we find that around noon, which is not incidentally precisely the area where the increase in load is extremely high (due to nonstationarity of the arrival stream), the prescribed number of servers follows the same slope as that of the square-root of the variance.
However, apparently this is not enough; the system can not deal with the backlog that is rapidly building up around noon.
Based on this observation, we cook up a heuristic that could potentially overcome this hurdle:
the hedge $\beta \sqrt{v_\infty(t)}$ is replaced by a more involved one, that accounts for extreme fluctuations in the arrival rate in settings where the level of nonstationarity is high.

\vspace{2mm}

\noindent \textit{Slope heuristic.} 
Let $v_n$ the ratio between the variance in time slot $n+1$ and $n$.
The idea is to scale up the number of servers when $v_n \gg 1$ while mildly reducing the number of servers when $v_n<1$, without changing the total number of staffed servers over the day.
It is important to only make subtle changes, so that the `shape' (viz., Figure \ref{fig:results0}) prescribed by the infinite-server proxy stays unaltered.
Consequently, we are after an increasing function $f(x)$ with the property that
\[
|f(x)-1| \leq |x-1|.
\]
Functions $f_\delta(x)=x^{\delta}$ with $0 < \delta \leq 1$ satisfy these conditions and have the advantage that they can easily be tuned via the parameter $\delta$.
We arrive (for $t=n\Delta$) at
\begin{equation}\label{superrule2}
s^m_{n\Delta}=m_\infty(n\Delta) + \beta \, (v_n)^{1/\delta}\sqrt{v_\infty(n\Delta)},
\end{equation}
for $n=1,\dots,24$.
Then $\delta$ can be picked such that the variance of the resulting delay probability (given a staffing level according to $s^\delta_{n\Delta}$ for $n=1,\dots,24$) is minimized. 
Alternatively, a few values for $\delta$ are compared to arrive at a value for which this variance is relatively small.
\hfill$\Diamond$

\begin{remark}\em 
Note that Eqn.\ \eqref{superrule2} simplifies to Eqn.\ \eqref{superrule} if $\lambda(t)$ is constant. 
The heuristic introduces an extension to the staffing rule that was originally proposed to account for nonstationarity; if there is no nonstationarity present ($\lambda(t)\equiv\lambda$), the slope-adapted rule reduces to the original rule, in which case the latter's performance is satisfactory.

Moreover, note that in the infinite-server setting performance would not improve by using the rule in Eqn.\ \eqref{superrule2}; in this setting Eqn.\ \eqref{superrule} is the best we can get.
That is to say, implementation of this heuristic is useful in situations where an extremely steep slope in the arrival rate causes an avalanche of queueing patients once the number of servers is restricted. 
\hfill$\Diamond$
\end{remark}

\begin{figure}[ht!]
\centering
\includegraphics[scale=0.597]{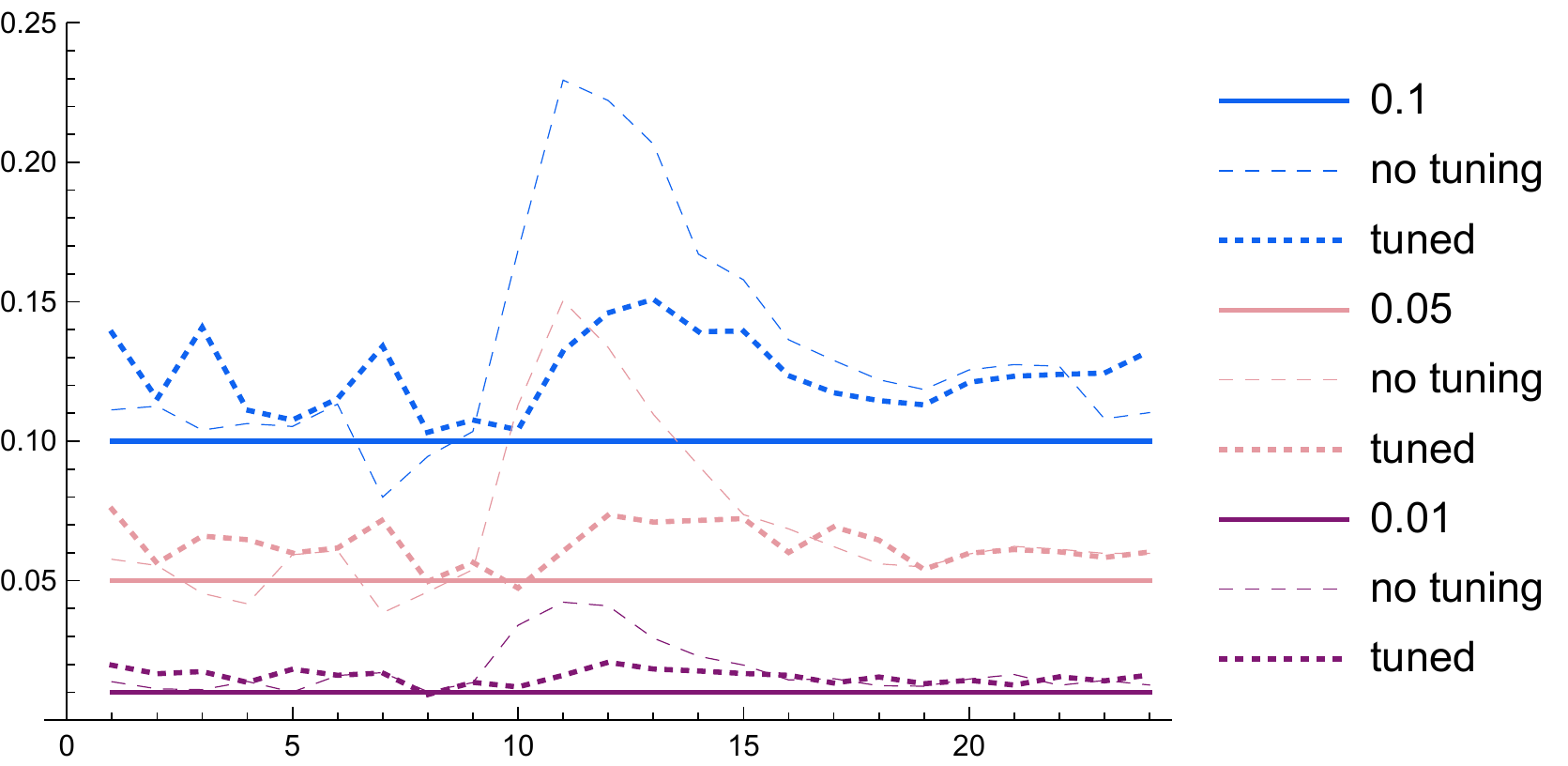}
\caption{\label{fig:results2}Using the slope-adapted staffing rule improves the stability of the delay probability. The delay probabilities with no tuning (dashed lines) coincide with those in Fig.\ \ref{fig:results1}. The dotted lines depict the delay probabilities after tuning and are clearly more stable than before. With $\delta_{\eps}$ denoting the selected $\delta$ for service level $\eps$, $\delta_{0.1}=3/8$, $\delta_{0.05}=1/3$ and $\delta_{0.01}=1/4$.}
\end{figure}

Figure \ref{fig:results2} compares the performance of the slope-adapted staffing rule (cf.\ Eqn.\ \eqref{superrule2}) with that of the originally proposed staffing rule (cf.\ Eqn.\ \eqref{superrule}). 
We observe better stability with (approximately) the same number of servers (in the example of Figure \ref{fig:results2} the total number of servers for both rules differs by 2 or 3 servers) and (on average) a slightly smaller delay probability. 
Nevertheless, the delay probabilities still exceed the targeted service level $\eps$.
Hence, the simple adaptation of choosing a higher value of $\beta$ in Eqn.\ \eqref{superrule} would further improve performance.

\section{Statistical procedures}\label{gap}

In this section we describe how to determine, based on historical data, the parameters in the arrival stream model introduced in Section \ref{sec2}, i.e., $\lambda_j$ for $j = 0, \dots, N-1$, $\alpha$, $I$, and $\Var W$.
Given a value $\Delta > 0$ it is enough to know, for each $j$:
$\bar{\Lambda}_j$, the average number of arrivals in 
$\Delta_j:=[j\Delta,(j+1)\Delta)$, and 
$\Sigma := \Sigma(\alpha,\Var W)$, the covariance matrix, representing for $j = 0, \dots, N-1$ 
and $k = 1, \dots, I$ the (nonnegative) covariance between the number of arrivals in $\Delta_j$ and $\Delta_{j+k}$ (note that here and in what follows, the indices in the subscripts should be taken modulo $N$; for the sake of readability we do not write that explicitly).

\vspace{2mm}

\noindent {\it Deterministic trend.}
The average number of arrivals $\bar{\Lambda}_j$ 
should correspond to the average over values of the mixed Poisson random variable with random parameter 
\[
\Lambda_j := \lambda_j \cdot c_{\alpha}\sum_{\ell = 0}^I \alpha^\ell W_{j-\ell},
\]
with $c_{\alpha} :=(1-\alpha)/(1-\alpha^{I+1})$. 
Using that $\E \Lambda_j = \lambda_j$, the $W_j$ have unit mean and that $c_\alpha$ is a normalizing constant, the $\bar{\Lambda}_j$ are unbiased estimators for the $\lambda_j$.

\vspace{2mm}

\noindent {\it Covariance matrix.}
Recall that the covariance of two independent mixed Poisson random variables (meaning that the enveloping Poisson random variables are independent) with dependent parameters equals the covariance of the parameters.
In addition, recall that the variance of a mixed Poisson random variable is the sum of the expectation and variance of its parameter.

Given that $I \leq \lfloor (N-1)/2 \rfloor$ (cf.\ Appendix \ref{app3}), we obtain the following expressions for the entries of the covariance matrix: 

\begin{align}\label{cova}
\Sigma_{j,j}=\E \Lambda_j + \Var{{\Lambda}_j}=\lambda_j + \lambda^2_j c_{\alpha}^2 \frac{1-\alpha^{2(I+1)}}{1-\alpha^2}\Var W;
\end{align}

\begin{align}
\nonumber
\Sigma_{j,j+k}=\Sigma_{j+k,j}&=\Cov({\Lambda}_j,{\Lambda}_{j+k})\\
&=\lambda_j \lambda_{j+k}  c_{\alpha}^2 \Cov\left(\sum_{\ell = 0}^I \alpha^\ell W_{j-\ell},\sum_{\ell = 0}^I \alpha^\ell W_{j+k-\ell}\right)\label{covb} \\
&=\lambda_j \lambda_{j+k}  c_{\alpha}^2\alpha^k \frac{1-\alpha^{2(I-k+1)}}{1-\alpha^2}\Var W, \label{covc}
\end{align}
where it's noted that $\Sigma_{j,j+k}=\Sigma_{j+k,j}=0$ for $k > I$. 
Let $C_k(\alpha,I) := c_{\alpha}^2\alpha^k \frac{1-\alpha^{2(I-k+1)}}{1-\alpha^2}$.
Then Eqns.\ \eqref{cova} and \eqref{covc} can be captured by 
\begin{equation}\label{covd}
\Sigma_{j,j+k} = \Sigma_{j+k,j} = \lambda_j \left( {{\rm \bf 1}}_{\{k=0\}} + \lambda_{j+k} C_k(\alpha,I) \Var W \right),
\end{equation}
for $k=0,1,\dots,I$ (and $0$ otherwise).
Note that by l'H\^opital's rule
\begin{equation*}
\lim_{\alpha \uparrow 1} C_k(\alpha,I) = \lim_{\alpha \uparrow 1} c_{\alpha}^2\alpha^k \frac{1-\alpha^{2(I-k+1)}}{1-\alpha^2} = \frac{I-k+1}{(I+1)^2}.
\end{equation*}
Hence we set $C_k(1,I):=(I-k+1)/(I+1)^2$.

\vspace{2mm}

\noindent {\it Procedure for $\alpha$, $I$ and $\Var W$.}
The idea is to vary $I$ in an outer loop and to estimate $\alpha$ and $\Var W$ (for any given $I$); one could then compare how much gain is made by using different $I$ with respect to the base case where $I=\Var W = 0$ (standard Poisson) and $I=0$ (no correlation).
Subsequently, it makes sense to select the largest $I$ that is a significant improvement over $I-1$ (or over the standard Poisson case, where $I=\Var W=0$). 
Note that the model only allows for values of $I$ ranging from $0$ to $11$. 

\begin{table}[ht!]
\centering
\begin{tabular}{|r|rrrr|r|}
\hline
$I$     & $\alpha$ & $\Var W$ & MSE* & MSE & Gain (\%) \bigstrut[t]\\
\hline
Poisson & -     & -     & -     & 7.435 & 0.000 \% \bigstrut[t]\\
\hline
0     & -        & 0.017 & 5.562 & 5.974 & 19.645 \% \bigstrut[t]\\
1     & 1.000 & 0.041 & 5.207 & 4.343 & 41.583 \% \\ 
2     & 1.000 & 0.060 & 4.267 & 3.476 & 53.244 \% \\ 
3     & 1.000 & 0.075 & 3.389 & 2.998 & 59.681 \% \\ 
4     & 1.000 & 0.089 & 2.821 & 2.685 & 63.881 \% \\ 
5     & 1.000 & 0.102 & 2.569 & 2.463 & 66.872 \% \\ 
6     & 0.907 & 0.112 & 2.344 & 2.342 & 68.501 \% \\ 
7     & 0.879 & 0.121 & 2.182 & 2.231 & 69.992 \% \\ 
8     & 0.866 & 0.129 & 2.246 & 2.120 & 71.481 \% \\ 
9     & 0.867 & 0.138 & 2.234 & 2.015 & 72.905 \% \\ 
10   & 0.866 & 0.146 & 2.098 & 1.936 & 73.963 \% \\ 11   & 0.861 & 0.152 & 1.949 & 1.894 & 74.520 \% \bigstrut[b]\\ 
\hline
\end{tabular}
\caption{Fitted parameters for Wednesday, Hospital 3.}
\label{table:gain}
\end{table}

\begin{table}
\centering
\begin{tabular}{|r|rrrr|r|}
\hline
$I$     & $\alpha$ & $\Var W$ & MSE* & MSE & Gain (\%) \bigstrut[t]\\
\hline
Poisson & -     & -     & -     & 12.253 & 0.000 \% \bigstrut[t]\\
\hline
0     & -         & 0.015 & 9.818 & 9.752 & 20.4 \% \bigstrut[t]\\
1     & 1.00 & 0.034 & 11.212 & 7.485 & 38.9 \% \\ 
5     & 1.00 & 0.084 & 6.238 & 4.566 & 62.7 \% \\ 
10   & 0.81 & 0.11 & 4.414 & 4.158 & 66.1 \% \\ 
\hline
\end{tabular}
\caption{Fitted parameters for Sunday, Hospital 3.}
\label{table:gainSun}
\end{table}

To be able to determine the values of $\alpha$ and $\Var W$ given $\lambda_j$, $\lambda_{j+k}$ and $I$, we need the empirical covariance matrix $\Sigma$ derived from the arrival data.
Note that an estimate of any two nonzero entries of $\Sigma$ provides enough information to solve for $\alpha$ and $\Var W$, after having equated them to the expression in Eqn.\ \eqref{covd}.
However, each nonzero pair leads to a different solution. 
We wish to determine values for $\alpha$ and $\Var W$ such that the theoretical covariance matrix $\Sigma(\alpha,\Var W)$ as given by Eqn.\ \eqref{covd} is the `best' approximation for $\Sigma$.
Therefore, the next step in the procedure is to minimize the average of the entrywise mean squared errors, where we sum over the entries for which the theoretical covariance matrix is nonzero (noting that the number of nonzero entries, being equal to $N(2I+1)$, depends on the choice of $I$).
In Table \ref{table:gain} this value is labeled with MSE*, with a separate column for the exact MSE values where all entries of the empirical covariance matrix are taken into account.
The gain is computed as the relative gain in (exact) MSE compared to the standard Poisson case (where $I = \Var W = 0$).
We observe that from $I=5$, not much improvement is still to be gained, so $I=5$ seems to be a good choice when we aim for moderate complexity and a good fit.

\vspace{2mm}

\noindent {\it Some more examples.}
In Table \ref{table:gainSun} the same procedure is used to obtain the best fit for $I=0,1,5,10$, which are used in the case study in Section \ref{intro.data}.
Similar gain percentages in MSE are obtained by choosing $I$ larger, although in Table \ref{table:gain} $I=5$ and $I=10$ achieve a better fit than in Table \ref{table:gainSun}.
At the same time, the variance of $W$ as well as the correlation parameter $\alpha$ is consistently smaller in the data set that corresponds to Sundays; although total arrival volume is larger here, temporal correlation and overdispersion seems to be less prominent.  

\begin{table}[ht!]
\centering
\begin{tabular}{|r|rr|r|}
\hline
$I$     & $\alpha$ & $\Var W$ & MSE\\
\hline
Poisson & - & - & 0.00 \\
\hline
0     & 1.00 & 0.172 & 87.8\\
1     & 0.871 & 0.343 & 18.5 \\ 
2     & 0.561 & 0.425 &3.99 \\ 
3     & 0.518 & 0.467 & 0.775 \\ 
4     & 0.505 & 0.489 & 0.104 \\ 
\rowcolor{cyan}
5     & 0.500 & 0.500 &0 \\ 
6     & 0.499 & 0.507 & 0.0240 \\
7     & 0.498 & 0.510 & 0.0415 \\ 
8     & 0.496 & 0.510 & 0.0483 \\
\hline
\end{tabular}
\caption{Fitted parameters given $I$ and corresponding MSE.}
\label{table:parameterfit}
\end{table}

We add a theoretical example to assess the precision of this minimization method.
With the $\lambda_j$ as above, set $\alpha^*=\Var W^* = 0.5$ and $I^*=5$.
As these parameters together define the arrival process, this gives a certain covariance matrix $\mathrm{Cov}^*$.
We use the minimization method to find, given some choice of $I \in \{0,1,\dots,7\}$, the optimal values for $\alpha$ and $\Var W$ in terms of the MSE of $\mathrm{Cov}(\alpha, \Var W)$ with respect to $\mathrm{Cov}^*$.
The results can be found in Table \ref{table:parameterfit}, together with the corresponding MSE.
Observe that the method recovers the true values $\alpha^*$ and $\Var W^*$ in case we set $I=I^*=5$.
For lower degree of correlation, it is found that a larger value for $\alpha$ (more dependence) is compensated by a smaller value for $\Var W$ (less overdispersion), however apart from the case $I=4$ choosing $I$ too small inevitably leads to a big loss in precision.
For $I=4$ the MSE is acceptably small.
On the other hand, setting $I$ too large leads only to small errors, which means that selecting a value $I$ above the true value leads to marginal differences.

\section{Performance}\label{per}

In this section we extend the numerical work on the performance of the presented staffing rules in the finite-server setting with described arrival stream, based on simulations instead of data.
The goal is to assess the individual effects of nonstationarity, temporal correlation and overdispersion. 
Moreover, to reduce the gap between the finite-server system and its infinite-server proxy we add an abandonment rate $\theta$ to the model.  
Note that in any service system that involves waiting, it is natural to have a (possibly small) positive abandonment rate.

We start with a particular stylized instance for the arrival stream, again inspired by the hospital data, i.e., the levels of overdispersion, nonstationarity and temporal correlation are comparable. 
The daily pattern is represented by a sine function with a cycle length of $24$ hours (with $\Delta=1$ hour), having a dip early in the morning (at 4:30) and a peak late in the afternoon (at 16:30).
That is, 
\[ 
\lambda_j = N + p N \cdot \sin(\frac{2\pi}{24}(j + 13.5)) \quad \text{ for } j=0,\dots, 23,
\]
where $N$ is the system size and $p$ the level of nonstationarity.
The parameters are set as in Table \ref{table:pars}.
\begin{table}[ht!]
\centering
\begin{tabular}{ | >{$} r <{$} @{\;=\;} >{$} l <{$} | >{$} r <{$} @{\;=\;} >{$} l <{$} |}
\hline
N & 17.5 & I & 5 \\
p & 0.8 & \alpha & 1 \\
\mu & 0.5 & \Var W & 0.1\\
\hline
\end{tabular}
\caption{Parameter setting base case.}
\label{table:pars}
\end{table}

Note that, as in Section \ref{intro.data}, the concerning system is fairly small whereas the level of nonstationarity is extremely high. 
As the service rate is low, this means that the patient `sees' effectively different arrival rates during its stay and nonstationarity can not be ignored.
The correlation structure is abundantly present and the level of overdispersion seems mild (though nonzero).
However, the effect of $\Var W$ being positive on the size of the hedge (i.e., on $\sqrt{v_{\infty}}$) is quite large; compare columns 1 and 3 in Table $\ref{table:ST1}$. 
On the other hand, taking $I>0$ slightly mitigates this effect; compare columns 2 and 3 in Table $\ref{table:ST1}$.
This table was included to show the effects of different modeling choices made independent of the impact of a (wildly fluctuating) daily pattern, which will be added in the experiment that follows.

In Table \ref{table:ST1} the staffing levels for $3 \times 3$ different settings are given.
Here abandonments are incorporated to different extents: the abandonment rate is $\theta = a*\mu$, for $a = 0, 0.5, 1$ ('no', 'mild' or 'max'). 
Note that incorporating abandonments does not affect the prescription for the staffing level, as our staffing rule does not account for it.  

We find that in this instance with stationary deterministic daily pattern, performance does not change significantly when a correlation structure is added to the model, where it is noted that we account for it in the staffing rule (in this case that means that less servers were used to achieve approximately the same delay probability).
We do however need significantly more servers to attain a comparable level for the delay probability when switching from the `no overdispersion' setting (column 1) to the setting with overdispersion (column 3), which of course is the motivation for the staffing rule introduced in this paper.
Note that performance gets worse anyway, despite the complication in the number of servers. 

From Table \ref{table:ST1} it becomes very clear that firm improvement in performance can be achieved by incorporating a positive abandonment rate. 
Nevertheless, we see that even without a daily pattern, the $\beta$ constant needs to be tuned somewhat until the delay probabilities match the targeted probabilities, partly due to the overdispersion (the first column displays better performance) and partly due to the inaccuracy of the infinite-server proxy (when abandonments are incorporated performance gets better until it's nearly perfect).
Strangely, only in the cases where $\eps = 0.1$ with no/mild abandonments, the performance got worse when correlation was left out.
Of course the proxy is least accurate in this case, but apparently having dependence between arrival rates `helps' here. 

{\small
\begin{table}[h]
\begin{tabular}{ccl|c|c|c|c|c|c|}
\cline{4-9}
&
& 
&
& standard
&
& $I>0$
&
& $I=0$
\\
\hline
& 
& $\sqrt{v_{\infty}}$
& $s$
& $5.91$
& $s$
& $7.02$
& $s$
& $8.06$
\\
\hline 
& 
& no abandonments
& 
& 0.098
& 
& 0.12
&
& 0.13
\\
& $\eps = 0.1$
& mild abandonments
& 44
& 0.086
& 45
& 0.10
& 46
& 0.11
\\
& 
& max abandonments
&
& 0.079
&
& 0.093
&
& 0.10
\\
\hline 
& 
& no abandonments
& 
& 0.051
&
& 0.055
&
& 0.068
\\
& $\eps = 0.05$
& mild abandonments
& 46
& 0.046
& 48
& 0.048
& 49
& 0.060
\\
& 
& max abandonments
&
& 0.043
&
& 0.044
&
& 0.056
\\
\hline 
& 
& no abandonments
&
& 0.011
&
& 0.017
&
& 0.016
\\
& $\eps = 0.01$
& mild abandonments
& 50
& 0.010
& 52
& 0.015
& 55
& 0.015
\\
& 
& max abandonments
& 
& 0.0099
&
& 0.014
&
& 0.014
\\
\hline
\end{tabular}
\caption{\label{table:ST1} Delay probability obtained through simulation, for the setting without nonstationarity (we set $p=0$). The first column gives the probability when setting $\Var W = 0$, in the third column the correlation structure is ignored ($I=0$). The middle column is the delay probability as dictated by our model given the stated parameter setting (see Table \ref{table:pars}).}
\end{table}}

\begin{figure}[htb]
	\begin{minipage}[t]{.3\textwidth}
		\centering
		\includegraphics[width=\textwidth]{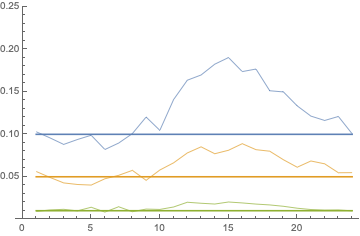} 
		\subcaption{\,}\label{fig:1}
	\end{minipage}
	\hfill
	\begin{minipage}[t]{.3\textwidth}
		\centering
		\includegraphics[width=\textwidth]{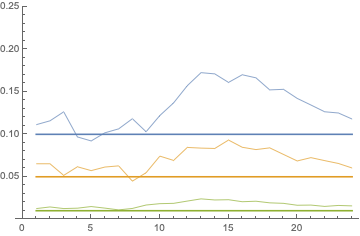} 
		\subcaption{\,}\label{fig:2}
	\end{minipage}
	\hfill
	\begin{minipage}[t]{.3\textwidth}
		\centering
		\includegraphics[width=\textwidth]{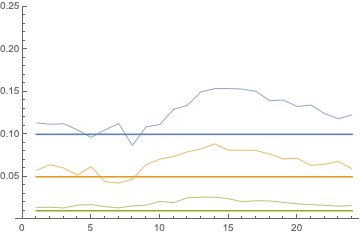}
		\subcaption{\,}\label{fig:3}
	\end{minipage}
	\hfill
	\begin{minipage}[t]{.3\textwidth}
		\centering
		\includegraphics[width=\textwidth]{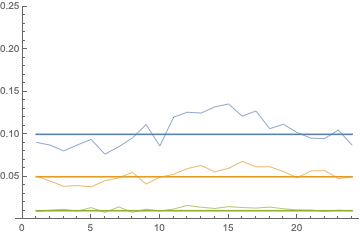} 
		\subcaption{\,}\label{fig:4}
	\end{minipage}
	\hfill
	\begin{minipage}[t]{.3\textwidth}
		\centering
		\includegraphics[width=\textwidth]{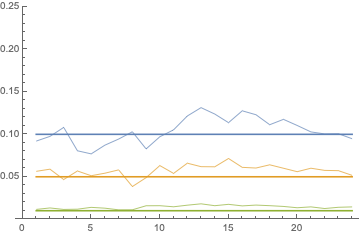} 
		\subcaption{\,}\label{fig:5}
	\end{minipage}
	\hfill
	\begin{minipage}[t]{.3\textwidth}
		\centering
		\includegraphics[width=\textwidth]{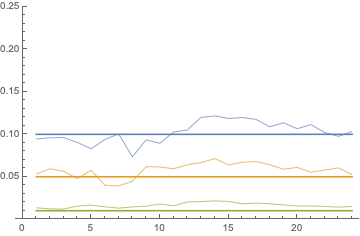}
		\subcaption{\,}\label{fig:6}
	\end{minipage}
	\hfill
	\begin{minipage}[t]{.3\textwidth}
		\centering
		\includegraphics[width=\textwidth]{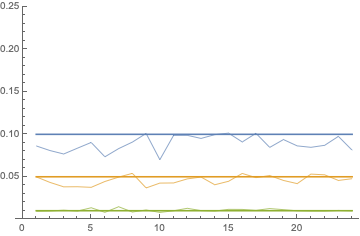} 
		\subcaption{\,}\label{fig:7}
	\end{minipage}
	\hfill
	\begin{minipage}[t]{.3\textwidth}
		\centering
		\includegraphics[width=\textwidth]{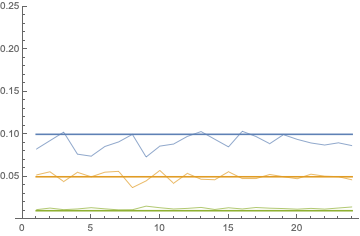} 
		\subcaption{\,}\label{fig:8}
	\end{minipage}
	\hfill
	\begin{minipage}[t]{.3\textwidth}
		\centering
		\includegraphics[width=\textwidth]{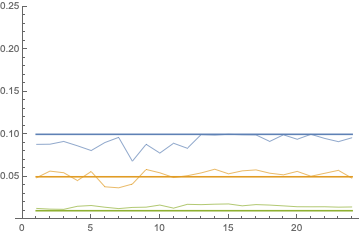}
		\subcaption{\,}\label{fig:9}
	\end{minipage}
	\hfill
\caption{\label{fig:CASES} Delay probability in finite-server setting, staffing level according new rule.}
\end{figure} 

From the plots in Figure \ref{fig:CASES} we observe that incorporating only mild abandonments significantly improves the performance of our staffing rule, which makes sense as the infinite-server proxy is more accurate for finite-server models with abandonments.
Note that the finite-server setting endowed with an abandonment rate $\theta = \mu$ coincides with the infinite-server setting, the setting in the last row of plots.
Note that the somewhat erratic nature
of the delay probability is due to inevitable rounding errors resulting from the fact that the number of servers needs to be integer.

In the first column we set $\Var W$ to zero (Subfigures (a), (d) and (g)), in the arrival stream as well as in the staffing rule.

Given that the delay probabilities in this setting define some sort of baseline for the performance (this should be the easiest setting to handle), it is remarkable that the performance does not get (significantly) worse when taking into account overdispersion and correlation.
In that sense it looks like our staffing rule is prescribing the correct number of servers.
The plots even suggest slight improvement in many settings.
Comparing the overdispersed setting where correlation is left out (column 3) with the setting with both overdispersion and correlation (column 2), there is only a very slight improvement in performance over all plots with different abandonment rates and staffing levels. 
That is, our rule seems to account well for overdispersion, but it is struggling slightly harder to deal with the correlation structure.
However, it can be concluded that nonstationarity is the main factor that complicates achieving stable delay probabilities, mostly in the setting without abandonments. 

\begin{figure}[htb]
	\begin{minipage}[t]{.3\textwidth}
		\centering
		\includegraphics[width=\textwidth]{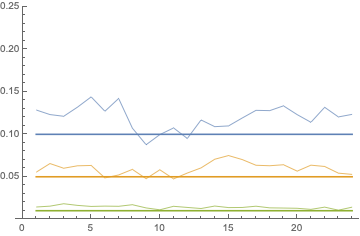} 
		\subcaption{\,}\label{fig:1b}
	\end{minipage}
	\hfill
	\begin{minipage}[t]{.3\textwidth}
		\centering
		\includegraphics[width=\textwidth]{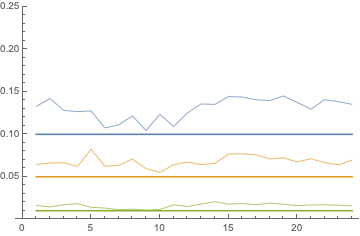} 		
		\subcaption{\,}\label{fig:2b}
	\end{minipage}
	\hfill
	\begin{minipage}[t]{.3\textwidth}
		\centering
		\includegraphics[width=\textwidth]{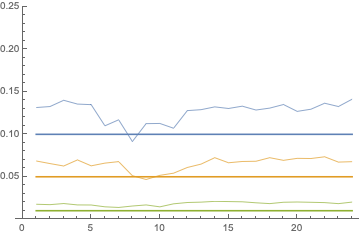} 
		\subcaption{\,}\label{fig:3b}
	\end{minipage}
	\hfill
\caption{ Delay probability in finite-server setting, staffing level determined with slope-adapted staffing rule. Here $\delta$ from Eqn.\ \eqref{superrule2} is of the form $\delta=\frac{k}{24}$, for the value of $k$ that maximally stabilizes the delay probability.} \label{fig:CASES2}
\end{figure} 

In order to make a fair comparison with the case study in Section~\ref{intro.data}, it is necessary to apply the slope-adapted staffing rule (cf.\ Eqn.\ \eqref{superrule2}) here as well.
We will only apply it to the case with no abandonments, hence we mirror Figure~\ref{fig:CASES} (a), (b) and (c): see Figure~\ref{fig:CASES2} (a), (b) and (c).
From Figure~\ref{fig:CASES2} it can be concluded that the slope-adapted staffing rule indeed stabilizes the delay probabilities over the day. 
However, further improvement could be made by tweaking $\beta$, to get the stabilized probabilities below the targeted level.
The resulting improvement is not shown, as the procedure and its effect are trivial.
                                                                                                                                                                                                                                                                 
\section{Conclusion and discussion}\label{CD}

In this paper we propose new staffing rules for a specific queueing model with overdispersed and nonstationary input with temporal correlation.
The objective is to stabilize the delay probability throughout the day around a fixed target value, which the final staffing rule developed succeeds to do.

In the numerical experiments in Section \ref{intro.data} the originally proposed rule based on an infinite-server proxy proves insufficient for staffing purposes.
The main complication turns out to be nonstationarity.
Considering the same model with abandonments, we observe significantly better performance, due to the fact that the infinite-server proxy is more accurate for finite-server models with abandonments.
Applying the introduced slope heuristic, the adapted staffing rule renders a major improvement (already without abandonments!).

The observed performance is robust for the choice of parameters for overdispersion and temporal correlation; as long as the combination of parameters results in an accurate estimate for the variance in the number of arriving customers, the prescribed (slope-adapted) staffing level is appropriate.
In Appendix \ref{app2}, it is shown that this variance is decreasing in $\alpha$ and $I$ and at the same time increasing in $\Var W$, so that different parameter settings can result in the same variance.
Although the statistical procedure in Section \ref{gap} does not lead to a unique `optimal' choice for the parameters $\alpha$, $I$ and $\Var W$, because of this robustness it is sufficient to select a reasonable parameter setting.

Note that the implementation of nonstationarity in the model is rather straightforward: fitting a constant arrival rate to fixed time slots is the simplest and also a widely used procedure to implement time-of-day or time-of-week effects. 
It is remarked, though, that the discontinuities might cause poor predictions close to the slot boundaries.
Therefore, in \cite{ZG17} a slight adaptation is suggested: it is proposed to use piecewise linear (hence continuous) rates.

\subsection*{Acknowledgments}
The authors would like to thank Johan van Leeuwaarden (Tilburg University) for useful suggestions and advice.


\bibliography{ReferencesMH}

\begin{thebibliography}{10}

\bibitem{Aksin2007}
Z.~Aksin, M.~Armony, and V.~Mehtotra.
\newblock The modern call center: A multi-disciplinary perspective on
  operations management research.
\newblock {\em Production and Operations Management}, 16(6):665--688, 2007.

\bibitem{ADE04}
A.N. Avramidis, A.~Deslauriers, and P.~L'Ecuyer.
\newblock Modeling daily arrivals to a telephone call center.
\newblock {\em Management Science}, 50(7):896--908, 2004.

\bibitem{BRZ10}
A.~Bassamboo, R.S. Randhawa, and A.~Zeevi.
\newblock Capacity sizing under parameter uncertainty: safety staffing
  principles revisited.
\newblock {\em Management Science}, 56(10):1668--1686, 2010.

\bibitem{BZ09}
A.~Bassamboo and A.~Zeevi.
\newblock On a data-driven method for staffing large call centers.
\newblock {\em Operations Research}, 57(3):714--726, 2009.

\bibitem{Borst2004}
S.C. Borst, A.~Mandelbaum, and M.I. Reiman.
\newblock Dimensioning large call centers.
\newblock {\em Operations Research}, 52(1):17--34, 2004.

\bibitem{CH01}
B.P.K. Chen and S.G. Henderson.
\newblock Two issues in setting call center staffing levels.
\newblock {\em Annals of Operations Research}, 108(1):175--192, 2001.

\bibitem{GKM03}
N.~Gans, G.~Koole, and A.~Mandelbaum.
\newblock Telephone call centers: Tutorial, review, and research prospects.
\newblock {\em Manufacturing \& Service Operations Management}, 5(2):79--141,
  2003.

\bibitem{Grassmann88}
W.K. Grassmann.
\newblock Finding the right number of servers in real-world queuing systems.
\newblock {\em Interfaces}, 18(2):94--104, 1988.

\bibitem{GKW07}
L.~V. Green, P.~J. Kolesar, and W.~Whitt.
\newblock Coping with time-varying demand when setting staffing requirements
  for a service system.
\newblock {\em Production and Operations Management}, 16(1):13--39, 2007.

\bibitem{GK91}
L.V. Green and P.~Kolesar.
\newblock The pointwise stationary approximation for queues with nonstationary
  arrivals.
\newblock {\em Management Science}, 37(1):84--97, 1991.

\bibitem{GKS91}
L.V. Green, P.~Kolesar, and A.~Svoronos.
\newblock Some effects of nonstationarity on multiserver {M}arkovian queueing
  systems.
\newblock {\em Operations Research}, 39(3):502--511, 1991.

\bibitem{GLT10}
I.~Gurvich, J.~Luedtke, and T.~Tezcan.
\newblock Staffing call-centers with uncertain demand forecasts: a
  chance-constrained optimization approach.
\newblock {\em Management science}, 56(7):1093--1115, 2010.

\bibitem{HW81}
S.~Halfin and W.~Whitt.
\newblock Heavy-traffic limits for queues with many exponential servers.
\newblock {\em Operations Research}, 29(3):567--588, 1981.

\bibitem{HLW16}
B.~He, Y.~Liu, and W.~Whitt.
\newblock Staffing a service system with non-{P}oisson nonstationary arrivals.
\newblock {\em Probability in the Engineering and Informational Sciences},
  30(4):593--621, 2016.

\bibitem{HLM17}
M.~Heemskerk, J.S.H. van Leeuwaarden, and M.~Mandjes.
\newblock Scaling limits for infinite-server systems in a random environment.
\newblock {\em Stochastic Systems}, 7(1):1--31, 2017.

\bibitem{IERS12}
R.~Ibrahim, P.~L'Ecuyer, N.~Regnard, and H.~Shen.
\newblock On the modeling and forecasting call center arrivals.
\newblock In {\em Proceedings of the 2012 Winter Simulation Conference}, pages
  1--12, 2012.

\bibitem{IYES16}
R.~Ibrahim, H.~Ye, P.~L'Ecuyer, and H.~Shen.
\newblock Modeling and forecasting call center arrivals: a literature survey
  and a case study.
\newblock {\em International Journal of Forecasting}, 32(3):865--874, 2016.

\bibitem{JLZ11}
A.J.E.M Janssen, J.S.H. van Leeuwaarden, and A.P. Zwart.
\newblock Refining square-root safety staffing by expanding {E}rlang-{C}.
\newblock {\em Operations Research}, 59(6):1512--1522, 2011.

\bibitem{JMMW96}
O.B. Jennings, A.~Mandelbaum, W.A. Massey, and W.~Whitt.
\newblock Server staffing to meet time-varying demand.
\newblock {\em Management Science}, 42(10):1383--1394, 1996.

\bibitem{JK01}
G.~Jongbloed and G.~Koole.
\newblock Managing uncertainty in call centers using {P}oisson mixtures.
\newblock {\em Applied Stochastic Models in Business and Industry},
  17(4):307--318, 2001.

\bibitem{Jouini}
O.~Jouini and S.~Benjaafar.
\newblock Appointment scheduling with non-punctual arrivals.
\newblock {\em IFAC Proceedings Volumes}, 42(4):235--239, 2009.

\bibitem{KVWC15}
S.-H. Kim, V.~Vel, W.~Whitt, and W.C. Cha.
\newblock Poisson and non-{P}oisson properties in appointment-generated arrival
  processes: The case of an endocrinology clinic.
\newblock {\em Operations Research Letters}, 43(3):247--235, 2015.

\bibitem{KW14}
S.-H. Kim and W.~Whitt.
\newblock Are call center and hospital arrivals well modeled by nonhomogeneous
  {P}oisson processes?
\newblock {\em Manufacturing \& Service Operations Management}, 16(3):464--480,
  2014.

\bibitem{KW14a}
S.-H. Kim and W.~Whitt.
\newblock Choosing arrival process models for service systems: Tests of a
  nonhomogeneous {P}oisson process.
\newblock {\em Naval Research Logistics (NRL)}, 61(1):66--90, 2014.

\bibitem{KWC17}
S.-H. Kim, W.~Whitt, and W.C. Cha.
\newblock A data-driven model of an appointment-generated arrival process at an
  outpatient clinic.
\newblock {\em INFORMS Journal on Computing}, 30(1):181--199, 2017.

\bibitem{KAW15}
Y.L. Ko\c{c}aga, M.~Armony, and A.R. Ward.
\newblock Staffing call centers with uncertain arrival rate and co-sourcing.
\newblock {\em Production and Operations Management}, 24(7):1101--1117, 2015.

\bibitem{LKDJ12}
S.~Liao, G.~Koole, C.~van Delft, and O.~Jouini.
\newblock Staffing a call center with uncertain non-stationary arrival rate and
  flexibility.
\newblock {\em OR Spectrum}, 34(3):691--721, 2012.

\bibitem{M09}
S.~Maman.
\newblock Uncertainty in the demand of service: The case of call centers and
  emergency departments.
\newblock Master thesis, Technion - Israel Institute of Technology, Haifa,
  2009.

\bibitem{MJLZ17}
B.W.J. Mathijsen, A.J.E.M. Janssen, J.S.H. van Leeuwaarden, and A.P. Zwart.
\newblock Robust heavy-traffic approximations for service systems facing
  overdispersed demand.
\newblock {\em Queueing Systems}, 90(3-4):257--289, 2018.

\bibitem{Mehrotra2010}
V.~Mehrotra, O.~Ozl\"uk, and R.~Saltzmann.
\newblock Intelligent procedures for intra-day updating of call center agent
  schedules.
\newblock {\em Production and operations management}, 19(3):353--367, 2010.

\bibitem{Robbins2010}
T.R. Robbins, D.J. Medeiros, and T.P. Harrison.
\newblock Does the {E}rlang {C} model fit in real call centers?
\newblock In {\em Proceedings of the Winter Simulation Conference}, pages
  2853--2864. Winter Simulation Conference, 2010.

\bibitem{RO79}
M.H. Rothkopf and S.S. Oren.
\newblock A closure approximation for the nonstationary ${M/M/s}$ queue.
\newblock {\em Management Science}, 25(6):522--534, 1979.

\bibitem{SHM09}
S.G. Steckley, S.G. Henderson, and V.~Mehrotra.
\newblock Forecast errors in service systems.
\newblock {\em Probability in the Engineering and Informational Sciences},
  23(2):305--332, 2009.

\bibitem{Tan2012}
J.~Tan, H.~Feng, X.~Meng, and L.~Zhang.
\newblock Heavy-traffic analysis of cloud provisioning.
\newblock In {\em Proceedings of the 24th International Teletraffic Congress},
  pages 1--8, 2012.

\bibitem{surveyJB}
J.S.H. van Leeuwaarden, B.~Mathijsen, and B.~Zwart.
\newblock Economies-of-scale in many-server queueing systems: tutorial and
  partial review of the {QED} {H}alfin-{W}hitt heavy-traffic regime.
\newblock {\em SIAM Review}, 61(3):403--440, 2019.

\bibitem{Leeuwaarden2016}
J.S.H. van Leeuwaarden, B.W.J. Mathijsen, and F.~Sloothaak.
\newblock Cloud provisioning in the {QED} regime.
\newblock In {\em Proceedings of the 9th EAI International Conference on
  Performance Evaluation Methodologies and Tools}, pages 180--187, 2016.

\bibitem{Whitt91}
W.~Whitt.
\newblock The pointwise stationary approximation for ${M_t/M_t/s}$ queues is
  asymptotically correct as the rates increase.
\newblock {\em Management Science}, 37(3):307--314, 1991.

\bibitem{Whitt92}
W.~Whitt.
\newblock Understanding the efficiency of multi-server service systems.
\newblock {\em Management Science}, 38(5):708--723, 1992.

\bibitem{Whitt93}
W.~Whitt.
\newblock Approximations for the ${GI/G/m}$ queue.
\newblock {\em Production and Operations Management}, 2(2):114--161, 1993.

\bibitem{Whitt99}
W.~Whitt.
\newblock Dynamic staffing in a telephone call center aiming to immediately
  answer all calls.
\newblock {\em Operations Research Letters}, 24(5):205--212, 1999.

\bibitem{Zan2012}
J.~Zan.
\newblock {\em Staffing service centers under arrival-rate uncertainty}.
\newblock PhD thesis, University of Texas, 2012.

\bibitem{ZLZ12}
B.~Zhang, J.S.H. van Leeuwaarden, and B.~Zwart.
\newblock Staffing call centers with impatient customers: refinements to
  many-server asymptotics.
\newblock {\em Operations Research}, 60(2):461--474, 2012.

\bibitem{ZG17}
Z.~Zheng and P.W. Glynn.
\newblock Fitting continuous piecewise linear {P}oisson intensities via maximum
  likelihood and least squares.
\newblock In {\em Proceedings of the 2017 Winter Simulation Conference}, pages
  1740--1749, 2017.

\end{thebibliography}
\bibliographystyle{plain}


\begin{appendix}
\section{Computations for infinite-server queue} \label{app1}
In this appendix we calculate $m_{\infty}(t)$ and $v_{\infty}(t)$ in terms of $\lambda(t)$, $\alpha$ and $\Var W$, which can be extracted from arrival data by means as proposed in Section \ref{gap}.
Let $\bar F(s):=\PR(S>s)$.
In this appendix we consider exponentially distributed service times, but
similar calculations can be done for other distributions in a straightforward manner.

Let $t=n \Delta$ for some $n \in \mathbb{Z}_{\geq 0}$.
We assume that $\lambda(s)$ is a periodic step function with step size $\Delta$ and cycle length $N$ (i.e., $\lambda(0)=\lambda(N \Delta)$) and write $\lambda_{k} := \lambda(t)$ for $t \in [k\Delta, (k+1)\Delta)$ for some non-negative value $\lambda_k$. 
As a consequence, for $\lambda_0, \dots, \lambda_{N-1}$, we have $\lambda_k = \lambda_\ell$ if $k\,{\rm mod}\, N=\ell\,{\rm mod}\, N.$

Let us start with evaluating $m_\infty(t)$ for this setting of periodic $\lambda(\cdot)$ and exponential service times (with mean $\mu^{-1}$). In the first place, an elementary calculation reveals that Eqn.\ \eqref{molmean} simplifies to (with $t =n\Delta$),
\begin{align*}
m_{\infty}(t) &= \E \left[\int_{0}^{\infty} \Lambda(t-u)\,\bar F(u)\, \mathrm{d}u\right]
= \frac{1-{\rm e}^{- \mu \Delta}}{\mu}\sum_{j=1}^{\infty} \lambda_{n-j}({\rm e}^{-\mu \Delta})^{j-1}
\end{align*}
For $j=1, \dots, N$, we introduce (using the periodicity)
\begin{align*}
\kappa_j(n) &:= \sum_{\ell=1}^{\infty} \lambda_{n-(\ell-1)N - j} ({\ee}^{-\mu \Delta})^{(\ell-1)N+j-1}= \lambda_{n-j}({\ee}^{-\mu \Delta})^{j-1}\sum_{\ell=1}^{\infty} ({\ee}^{-\mu \Delta N})^{\ell-1}= \lambda_{n-j} \cdot \frac{{\ee}^{-\mu \Delta(j-1)}}{1-{\ee}^{-\mu \Delta N}}.
\end{align*}
This leads to an expression for $m_{\infty}(t)$ in terms of a finite sum:
\begin{equation}\label{eqminfty}
m_{\infty}(t) 
= \frac{1-{\ee}^{- \mu \Delta}}{\mu} \sum_{j=1}^N \kappa_j(n)
= \frac{1-{\ee}^{- \mu \Delta}}{1-{\ee}^{-\mu \Delta N}}\frac{1}{\mu}\sum_{j=1}^N {\lambda_{n-j} }\,{\ee}^{-\mu \Delta(j-1)}.
\end{equation}

We now move on to compute $v_\infty(t)$. 
To this end, we define 
\[\gamma(j) := \lambda_{j} c_{\alpha} \int^{(n-j)\Delta}_{(n-j-1)\Delta}\bar F(u)\, \mathrm{d}u.
\]
The idea is to rearrange the contributions to the random arrival rate due to each of the $W_j$ in the expression for $v_\infty(t)$ in Eqn.\ \eqref{molvar}:
\begin{align*}
\Var{\int_{0}^{\infty} \Lambda(t-u)\,\bar F(u)\, \dd u}&= \Var {\sum_{j=1}^{\infty} \lambda_{n-j} \,\Big(c_{\alpha}\sum_{\ell = 0}^{I} \alpha^\ell W_{n-j-\ell}\Big) \int^{j\Delta}_{(j-1)\Delta}\bar F(u)\, \dd u}\\
& = \Var {\sum_{j=1}^{\infty} \,\Big(\lambda_{n-j} c_{\alpha} \int^{j\Delta}_{(j-1)\Delta}\bar F(u)\, \dd u \Big) \sum_{\ell = 0}^{I} \alpha^\ell  W_{n-j-\ell}}\\
& = \Var {\sum_{j=1}^{\infty} \gamma(n-j) \sum_{\ell = 0}^{I} \alpha^\ell  W_{n-j-\ell}}\\
& = \Var {\sum_{j=1}^{\infty} \bigg(\sum_{\ell = 0}^{I \land (j-1)} \alpha^\ell \gamma(n-j+\ell)\bigg) W_{n-j}}
\end{align*}
Noting that the $W_j$ are independent and identically distributed, the expression in the previous display becomes
\begin{align*}\Var W
 \sum_{j=1}^{I}& \bigg(\sum_{\ell = 0}^{j-1} \alpha^\ell \gamma(n-j+\ell)\bigg)^2  
+ \Var W\sum_{j=I+1}^{\infty} \bigg(\sum_{\ell = 0}^{I} \alpha^\ell \gamma(n-j+\ell)\bigg)^2    \\
& =\Var W c^2_{\alpha} \frac{({\ee}^{\mu \Delta}-1)^2}{\mu^2}  \, \sum_{j=1}^{\infty}\bigg(\sum_{\ell = 0}^{I \land (j-1)} \alpha^{\ell} \lambda_{n-j+\ell}{\ee}^{-\mu(j-\ell)\Delta}\bigg)^2,
\end{align*}
where we use that
\begin{align*}
\bigg(\sum_{\ell = 0}^{I \land (j-1)}& \alpha^\ell \gamma(n-j+\ell)\bigg)^2 = \left(\sum_{\ell = 0}^{I \land (j-1)} \alpha^\ell \lambda_{n-j+\ell} c_{\alpha}
\int^{(j-\ell)\Delta}_{(j-\ell-1)\Delta}\bar F(u)\, \mathrm{d}u \right)^2 
=
c^2_{\alpha} \cdot \frac{({\ee}^{\mu \Delta} - 1)^2}{\mu^2}B_{j}
\end{align*}
with
\[B_{j}:=
 \bigg(\sum_{\ell = 0}^{I \land (j-1)} \alpha^\ell \lambda_{n-j+\ell} ({\ee}^{-\mu\Delta})^{ j - \ell }\bigg)^2.\]
The next step is again to exploit the periodicity. 
For this we study $\sum_{j=1}^{\infty} B_{j}$, under the assumption $I < N$
(which is fairly natural).
Elementary calculus reveals that $v_{\infty}(t)$ can be expressed as a finite sum, due to
\begin{align*}
\sum_{j=1}^{\infty} &B_{j} = \sum_{j=1}^{\infty}\bigg(\sum_{\ell = 0}^{I \land (j-1)} \alpha^{\ell} \lambda_{n-(j-\ell)}({\rm e}^{-\mu \Delta})^{j-\ell}\bigg)^2 
= \sum_{j=1}^{\infty} \alpha^{2j} \bigg(\sum_{\ell = 0}^{I \land (j-1)}\lambda_{n-(j-\ell)}\left(\frac{{\ee}^{-\mu \Delta}}{ \alpha}\right)^{j-\ell}\bigg)^2\\
&= \sum_{j=1}^N\sum_{k=1}^{\infty} \alpha^{2j} \bigg(\sum_{\ell = 0}^{I \land (j-1)}\lambda_{n-(j-\ell) - (k-1)N}\left(\frac{{\ee}^{-\mu \Delta}}{ \alpha}\right)^{j-\ell + (k-1)N}\bigg)^2\\
&= \sum_{j=1}^N (\alpha^{2j} \bigg(\sum_{\ell = 0}^{I \land (j-1)}\lambda_{n-(j-\ell)}\left(\frac{{\ee}^{-\mu \Delta}}{ \alpha}\right)^{j-\ell}\bigg)^2 + \sum_{k=1}^{\infty} \alpha^{2j} \bigg(\sum_{\ell = 0}^{I}\lambda_{n-(j-\ell) - kN}\left(\frac{{\ee}^{-\mu \Delta}}{ \alpha}\right)^{j-\ell + k N}\bigg)^2 )\\
&= \sum_{j=1}^I \alpha^{2j} \bigg(\sum_{\ell = 0}^{(j-1)}\lambda_{n-(j-\ell)}\left(\frac{{\ee}^{-\mu \Delta}}{ \alpha}\right)^{j-\ell}\bigg)^2+ \sum_{j=I+1}^N \alpha^{2j}  \bigg(\sum_{\ell = 0}^{I}\lambda_{n-(j-\ell)}\left(\frac{{\ee}^{-\mu \Delta}}{ \alpha}\right)^{j-\ell}\bigg)^2 \\
&\quad \qquad \qquad + \sum_{j=1}^N \alpha^{2j}  \sum_{k=1}^{\infty} \left(\frac{{\ee}^{-\mu \Delta}}{ \alpha}\right)^{2kN}\bigg(\sum_{\ell = 0}^{I}\lambda_{n-(j-\ell)}\left(\frac{{\ee}^{-\mu \Delta}}{ \alpha}\right)^{j-\ell}\bigg)^2 \bigg)\\
&=\sum_{j=1}^I \alpha^{2j} \bigg(\sum_{\ell = 0}^{(j-1)}\lambda_{n-(j-\ell)}\left(\frac{{\ee}^{-\mu \Delta}}{ \alpha}\right)^{j-\ell}\bigg)^2  + \sum_{j=I+1}^N \alpha^{2j}  \bigg(\sum_{\ell = 0}^{I}\lambda_{n-(j-\ell)}\left(\frac{{\rm e}^{-\mu \Delta}}{ \alpha}\right)^{j-\ell}\bigg)^2 \\
&\quad \qquad \qquad + \frac{1}{\alpha^{2N}{\rm e}^{2\mu \Delta N}-1}\sum_{j=1}^N \alpha^{2j}  \bigg(\sum_{\ell = 0}^{I}\lambda_{n-(j-\ell)}\left(\frac{{\rm e}^{-\mu \Delta}}{ \alpha}\right)^{j-\ell}\bigg)^2 \bigg).
\end{align*}
Note that for convergence of the infinite series, we have to assume that ${\ee}^{-\mu \Delta} < \alpha$.

The final expression for $v_{\infty}(t)$ is as follows:
\begin{align}\label{eqvinfty}
v_{\infty}(t)
&=\Var W c^2_{\alpha} \frac{({\rm e}^{\mu \Delta}-1)^2}{\mu^2} 
\cdot
\sum_{j=1}^N \alpha^{2j} \cdot D_{j}, \\
\text{where } & D_{j} := \bigg(\sum_{\ell = 0}^{(j-1)\land I}\lambda_{n-(j-\ell)}\left(\frac{{\ee}^{-\mu \Delta}}{ \alpha}\right)^{j-\ell}\bigg)^2 
+ \frac{1}{\alpha^{2N}{\ee}^{2\mu \Delta N}-1} \bigg(\sum_{\ell = 0}^{I}\lambda_{n-(j-\ell)}\left(\frac{{\ee}^{-\mu \Delta}}{ \alpha}\right)^{j-\ell}\bigg)^2. \nonumber
\end{align}

\section{Variance of the arrival process as a function of the parameter space} \label{app2}

In this appendix we show that the variance of the arrival process is decreasing in $\alpha$ and $I$ and increasing in $\Var W$.
With the nonstationary doubly-stochastic arrival rate process $\Lambda(t)$ given by Eqn.\ \eqref{arr}, we get a nonstationary mixed Poisson arrival process. 
This process is overdispersed, which becomes visible once we write down its variance:
\begin{equation}\label{varr}
\Var{ \big( \mathrm{Poisson}(\Lambda(t))\big)}  = \lambda_j + \lambda_j^2 \frac{(1-\alpha)^2}{(1-\alpha^{I+1})^2} \frac{1-\alpha^{2(I+1)}}{1-\alpha^2}\Var{W},
\end{equation}
which is larger than $\lambda_j$ as the second term at the right-hand side of Eqn.\ \eqref{varr} is positive for $\Var{W}>0$.
It is hence directly seen that Eqn.\ \eqref{varr} is increasing in $\Var{W}$.
Observe that the factor 
\begin{equation}\label{alphaI}
\frac{(1-\alpha)^2}{(1-\alpha^{I+1})^2} \frac{1-\alpha^{2(I+1)}}{1-\alpha^2}=\frac{1-\alpha}{1+\alpha}\frac{1+\alpha^{I+1}}{1-\alpha^{I+1}}
\end{equation}
depends both on $\alpha$ and $I$.
We state and prove that Eqn.\ \eqref{alphaI} is (strictly) decreasing in $\alpha$ and $I$.
Note that indeed
\begin{equation*}
\frac{1+\alpha^{I+1}}{1-\alpha^{I+1}} < \frac{1+\alpha^{I}}{1-\alpha^{I}} \quad \text{ for }I =0,1,\dots, 
\end{equation*}
for all $\alpha \in (0,1)$.
Now consider the function $f_I(\alpha) = \frac{1-\alpha}{1+\alpha}\frac{1+\alpha^{I}}{1-\alpha^{I}}$ for some $I>1$ (for $I=1$ we find this function is constant; note that this corresponds to the case where $I=0$ in our model, i.e. there is no correlation between past time slots).
After taking the logarithm and differentiating, we end up with the condition that the function is (strictly) decreasing if
\[
I \alpha^{I-1}(\frac{1}{1+\alpha^I} + \frac{1}{1-\alpha^I}) < \frac{1}{1+\alpha} + \frac{1}{1-\alpha}.
\]
Rewriting gives
\[
I \alpha^{I-1} < \sum_{k=0}^{I-1} \alpha^{2k}=\begin{cases}
\alpha^{I-1} + \sum_{k=0}^{(I-1)/2-1} \left(\alpha^{2k}+\alpha^{2(I-1-k)}\right) &\text{ if $I$ is odd }\\
\sum_{k=0}^{I/2-1}\left( \alpha^{2k}+\alpha^{2(I-1-k)}\right) &\text{ if $I$ is even }\\
\end{cases},
\]
and these two cases are easy to check individually, since $\alpha^{I-1} < \frac12 \left(\alpha^{2k}+\alpha^{2(I-1-k)}\right)$ for all relevant $k$.

\section{Constraint on $I$}\label{app3}
In this appendix we explain why we take $I$, the number of elapsed time slots that affect the busyness factor, to be at most equal to $\lfloor (N-1)/2 \rfloor$.
Note that the number of nonzero entries in the covariance matrix equals $N(2I+1)$, 
as for each time slot $j=0,\dots,N-1$ we have a nonzero entry on the diagonal and for $k=1,\dots,I$ both $\Sigma_{j,j+k}$ and $\Sigma_{j,j-k}$ are nonzero.
Of course the dimension of the matrix only allows for $N^2$ entries.
In other words: $N(2I+1)\leq N^2$ must hold, i.e., 
\begin{equation}\label{restrict}
I \leq \lfloor (N-1)/2 \rfloor.
\end{equation}
To be even more precise, strictly it is only required to set $I \leq \lfloor N/2 \rfloor$, however in the case where $N$ is even, for $k=I=N/2$ we should replace Eqn.\ \eqref{covc} by
\begin{align}
\nonumber
\Sigma_{j,j+N/2}=\Sigma_{j+N/2,j}&=\Cov({\Lambda}_j,{\Lambda}_{j+N/2})\\
\nonumber
&=\lambda_j \lambda_{j+N/2}  c_{\alpha}^2 \Cov\left(\sum_{\ell = 0}^I \alpha^\ell W_{j-\ell},\sum_{\ell = 0}^I \alpha^\ell W_{j+k-\ell}\right)\\
&=\lambda_j \lambda_{j+N/2}  c_{\alpha}^2\alpha^{N/2}\Var W, \label{ex2b}  
\end{align}
for $j=0, \dots, N-1$.

The following example serves as an illustration of the complication that arises when $I$ is not restricted as in Eqn.\ \eqref{restrict}.
\begin{example}{\em 
Let $N=24$. 
Then Eqn.\ \eqref{covc} holds for $k = 1, \dots, 11$ (when we pick $I = 11$). 
If we were to choose $I = 12$ and used Eqn.\ \eqref{covb} for the covariance between arrivals in $\Delta_0$ and $\Delta_{12}$, we would get 
\begin{equation}\label{ex2}
\lambda_0 \lambda_{12} c_{\alpha}^2 \Cov\left(\alpha^I W_{12} + \dots + \alpha W_{23} + W_0,\alpha^{12} W_0 + \dots + \alpha W_{11} + W_{12}\right).
\end{equation}
However, the second occurence of $W_{12}$ in Eqn.\ \eqref{ex2} is incorrect and should be written as $W'_{12}$: it is describes an i.i.d.\ copy of $W_{12}$.
As a result, the covariance just equals $\lambda_0 \lambda_{12} c_{\alpha}^2 \alpha^{12} \Var {W_0}$ (cf.\ Eqn.\ \eqref{ex2b}).
Note that it's in fact still possible to write all nonnegative covariances in a $24 \times 24$-matrix.
Namely, for $I=11$ still $N$ entries equal zero; as for $I=12$ and $k=N/2=12$ the entries $\Sigma_{j,j+k}$ and $\Sigma_{j,j-k}$ happen to coincide (for $j=0,\dots,N-1$), these $N$ values exactly fill up the `previously unoccupied' entries.

If however, we had chosen $I=13$, we would need to write both the covariance 
between arrivals in $\Delta_0$ and $\Delta_{13}$ (where time slot $\Delta_0$ passes first) and the covariance between arrivals in $\Delta_{13}$ and the $\Delta_0$ after (i.e., $\Delta_{13}$ passes first) on the same entry, but their values do not match.
\hfill$\Diamond$}
\end{example}

All in all, we see that it makes sense to exclude (for simplicity) $I > \lfloor (N-1)/2 \rfloor$ from the parameter space, to ensure that the correlation in our model does not exceed intraday level. 

\end{appendix}

\end{document}